\documentclass[10pt]{iopart}
\usepackage{latexsym,epsfig,graphicx,bbm,amssymb}

\newcommand{\half}{\mbox{$\textstyle \frac{1}{2}$}}
\newcommand{\ket}[1]{\left | \, #1 \right \rangle}
\newcommand{\kets}[1]{| \, #1 \rangle}
\newcommand{\ceil}[1]{\lceil #1 \rceil}
\newcommand{\bra}[1]{\left \langle #1 \, \right |}

\newcommand{\av}[1]{\langle #1\rangle}
\newcommand{\outprod}[2]{\ket{#1}\bra{#2}}

\newcommand{\mod}{\,\textrm{mod}\,}

\newcommand{\eqr}[1]{Eq.~(\ref{#1})}
\newcommand{\fir}[1]{Fig.~\ref{#1}}
\newcommand{\secr}[1]{Sec.~\ref{#1}}

\begin{document}

\paper[Entanglement consumption of instantaneous nonlocal quantum
measurements]{Entanglement consumption of instantaneous nonlocal quantum
measurements}

\author{S R Clark$^{1,2}$, A J Connor$^{2}$, D Jaksch$^{2,1}$ and S Popescu$^{3}$}
\address{$^1$Centre for Quantum Technologies, National University of
Singapore, 3 Science Drive 2, Singapore 117543}
\address{$^2$Clarendon Laboratory, University of Oxford, Parks
Road, Oxford OX1 3PU, United Kingdom}
\address{$^3$ H.H. Wills Physics Laboratory, University of Bristol, Tyndall Avenue, Bristol BS8 1TL, United Kingdom}

\date{\today}
\ead{s.clark@physics.ox.ac.uk}

\begin{abstract}
Relativistic causality has dramatic consequences on the measurability of nonlocal variables and poses the fundamental question of whether it is physically meaningful to speak about the value of nonlocal variables at a particular time. Recent work has shown that by weakening the role of the measurement in preparing eigenstates of the variable it is in fact possible to measure all nonlocal observables instantaneously by exploiting entanglement. However, for these measurement schemes to succeed with certainty an infinite amount of entanglement must be distributed initially and all this entanglement is necessarily consumed. In this work we sharpen the characterisation of instantaneous nonlocal measurements by explicitly devising schemes in which only a finite amount of the initially distributed entanglement is ever utilised. This enables us to determine an upper bound to the average consumption for the most general cases of nonlocal measurements. This includes the tasks of state verification, where the measurement verifies if the system is in a given state, and verification measurements of a general set of eigenstates of an observable. Despite its finiteness the growth of entanglement consumption is found to display an extremely unfavourable exponential of an exponential scaling with either the number of qubits needed to contain the Schmidt rank of the target state or total number of qubits in the system for an operator measurement. This scaling is seen to be a consequence of the combination of the generic exponential scaling of unitary decompositions combined with the highly recursive structure of our scheme required to overcome the no-signalling constraint of relativistic causality.
\end{abstract}


\maketitle

\section{Introduction}
\label{intro}
The formal compatibility of quantum mechanics with special relativity is highly nontrivial~\cite{Peres} and is in many ways quite miraculous~\cite{Peskin}. Perhaps the most well known difficulty in combining these formalisms arises from the so-called ``collapse'' of a quantum state associated with the measurement process, and in particular the instantaneity of this change. This problem is highlighted in its most simple form by considering two observers, Alice and Bob, who are spacelike separated. Conventional wisdom holds that any self-adjoint operator that can be defined for Alice and Bob's joint system is measurable in principle~\cite{Dirac}. But in fact, most such operators represent nonlocal variables, meaning that they cannot be written in the form $A\otimes B$, where $A$ and $B$ are self-adjoint operators acting on Alice and Bob's local Hilbert spaces, respectively. Early on it was recognised that if such nonlocal variables were instantaneously measurable, in the standard sense in some Lorentz frame, then violations of relativistic causality arise (see Fig 1 for a description of this effect for an ideal measurements on two separated spin-$\half$ particles). In 1931 Landau and Peierls~\cite{Landau} claimed that this observation implied, quite generally, the impossibility of measuring any nonlocal variable at a well-defined time, and even went so far as to postulate a new uncertainty principle to this effect. Thus a common consensus arose that it only made sense to speak of local variables as observables in relativistic quantum mechanics.

It was only in 1980 that this conjecture was finally refuted by Aharonov and Albert~\cite{Aharonov80,Aharonov81} who explicitly constructed a scheme for measuring certain nonlocal variables (e.g. the Bell operator; see \secr{framework}) instantaneously without contradicting causality. In contrast to previous studies their measurement scheme explicitly introduced entangled probes whose quantum correlations enable nonlocal properties of the system to become correlated to local properties of the measuring device. This means that by combining the correlated local outcomes of the two observers, at some point in the future when their light cones have intersected, the final nonlocal measurement result can be revealed. Their discovery had serious implications for the notion of states and observables in relativistic quantum mechanics. It immediately disproved the previously held covariant state reduction postulate~\cite{Hellwig} whose validity was dependent on only local variables being measurable. It also showed that no covariant succession of states at a given time can be associated to the system since observers in different Lorentz frames will have conflicting accounts of the reduction process which cannot be reconciled within any single covariant state history. This far-reaching conclusion culminated in their proposing that to take account of changes to a state vector, induced by local or nonlocal measurement processes, it is required that the wavefunction ceases being a function of spacetime and instead becomes a functional on the set of spacelike hypersurfaces~\cite{Aharonov84a,Aharonov84b}.

Further work~\cite{Aharonov86} then detailed explicit methods for measuring nonlocal variables  such as $A + B$ and modular sums like $(A + B) \mod a$, where $a$ is a desired eigenvalue. It was later proven in generality by Popescu and Vaidman~\cite{Popescu} that any conceivable measurement requires the erasure of local information (within the relevant degrees of freedom) in order to be compatible with causality. For a standard non-demolition measurement, to satisfy this additional requirement there is a dramatic restriction on what is measurable. For the case of two spin-$\half$ particles causality limits the measurability of operators to those with either trivial direct product eigenstates or maximally entangled Bell states. The measurability of the latter is permitted because the reduced density matrix of either spin is always proportional to the unit matrix. A surprising consequence of this result is that even nonlocal variables with product eigenstates (see \eqr{twisted_basis} in \secr{twoqubit_measure}) are not measurable~\cite{Groisman01} showing further that standard quantum measurements can be non-separable in a way not entirely captured by the notion of entanglement\footnote{This situation is in contrast to the usual {\em nonlocality without entanglement} scenario where the constraint is on quantum resources and unlimited classical communication is assumed~\cite{Bennett99}.}. The information erasure theorem~\cite{Popescu} indicates that causal measurement schemes for almost all nonlocal variables cannot be a standard von-Neumann measurement. Such measurements leave the system undisturbed if it was in an eigenstate of the observable before the measurement and play a dual role of both observing a quantity and preparing the system in an eigenstate of the corresponding observable~\cite{Breuer, Nielsen}. It is now recognised that this framework, which was the basis of Landau and Peierls conjecture, is too restrictive to decide whether a nonlocal variable attains the status of a physical observable. Instead an operators measurability should be determined in a broader paradigm of verification measurements~\cite{Groisman,Vaidman03,Sorkin}. A verification measurement can confirm with certainty whether the system is in an eigenstate of an observable at a given time, but does not necessarily leave the system in an eigenstate after it is completed. These measurements are therefore destructive and non-repeatable. 

\begin{figure}[t]
\begin{center}
\includegraphics[width=13cm]{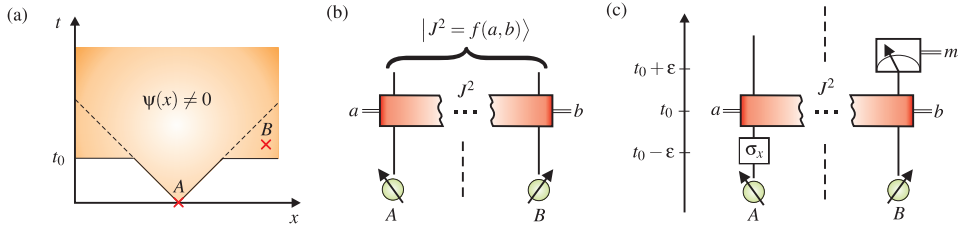}
\caption{(a) In a scenario envisaged by Landau and Peierls a particle is initially localized at a point $A$ and is subject to an ideal momentum measurement at some later time $t=t_0$. The effect of this measurement is to instantaneously collapse the particles wavefunction into a momentum eigenstate which subsequently redistributes the particles probability amplitude throughout all space. There is then a non-zero probability of finding the particle at a location $B$ which is spacelike separated from $A$. (b) A simpler formulation of this causality violation can be framed using spin-$\half$ particles (see \secr{framework}). Here we consider a device which can perform an instantaneous ideal measurement of the magnitude of the total spin squared $J^2=\sum_{k=x,y,z}(\sigma^A_k + \sigma^B_k)^2$ of two spacelike separated spins. Dependent on the local outcomes $a$ and $b$ via some function $f$ the state after the measurement will be projected into one with a well defined value $J^2 = f(a,b)$. (c) If such a measuring device exists it would violate relativistic causality. Suppose Alice and Bob prepare a state $\ket{\uparrow}_A\ket{\uparrow}_B$ in the distant past and arrange to measure $J^2$ at time $t_0$. If just prior to the $J^2$ measurement at time $t_0 - \epsilon$  Alice flips her spin then a measurement of $\sigma^B_z$ by Bob just after the $J^2$ measurement at time $t_0 + \epsilon$ will yield $\uparrow$ and $\downarrow$ with equal probability. Since the time interval $2\epsilon$ can be made arbitrarily small Alice can use the $J^2$ measurement to send superluminal signals.}\label{causality}
\end{center}
\end{figure}

Recent work in the context of gauge theories has further highlighted the fundamental implications of how measurability is defined~\cite{Beckman02}. In particular for gauge theories, which are used to describe all elementary particles, it is common to characterise gauge field configurations by Wilson loop operator's. These manifestly nonlocal quantities are taken to be basic observables in gauge theory. Yet it was shown that the non-demolition measurement of spacelike Wilson loops in a relativistic non-Abelian gauge theory violates causality, and that instead only verification measurements are possible~\cite{Beckman02}. From a different perspective it has also been shown recently that the use of additional ancillary resources can dramatically alter the properties of nonlocal measurements, for example by revealing Bell-inequality violations in delocalised single-particle mode entanglement that would otherwise be prohibited by super-selection rules~\cite{Ashhab,Heaney,Paterek}. Indeed by moving both to verification measurements and exploiting ancilla it has been found that there are no causal restrictions on what variables can be measured. Firstly, using methods devised for remote probabilistic rotations~\cite{Reznik} it was shown that all observables of two spin-$\half$ particles can be measured instantaneously~\cite{Groisman}. Secondly, a method based on teleportation~\cite{Bennett93} was devised which demonstrates the instantaneous measurability of all observables for multipartite systems of arbitrary dimension~\cite{Vaidman03, Groisman03}. These studies have therefore answered affirmatively that the instantaneous measurement of all nonlocal variables\footnote{One caveat to this, which applies to this work as well, are variables related to fermionic degrees of freedom that are spatially delocalised~\cite{Vaidman03,Aharonov00}.} can be achieved without contradicting quantum mechanics and causality. Thus in principle all nonlocal variables are valid physical observables and so within this framework the conventional wisdom is reestablished.

A critical ingredient in these measurement schemes is entanglement. However, since the main aim of those schemes~\cite{Groisman,Vaidman03} was disproving causal restrictions they were not concerned with limiting the amount of entanglement consumed. As a result to guarantee success in the most general cases these schemes require an unlimited supply of entanglement to be initially distributed between Alice and Bob, and all of this entanglement is necessarily consumed. Here we go beyond this by systematically addressing the latter issue, namely the entanglement consumption. Firstly we explicitly devise a scheme, which significantly optimizes that by Vaidman in~\cite{Vaidman03}, where only a finite amount of the initial entanglement is ever consumed on average. This enables us to sharpen the charaterization of instantaneous nonlocal measurements by quantifying the cost of nonlocal measurement tasks. Specifically we determine an upper bound to the average consumption for state verification, where the measurement verifies if the system is in a given state, and for the verification measurement of a general set of eigenstates of an observable. Secondly, it is straightforward to show from our scheme that by only allowing a finite amount of the initial entanglement, in addition to a finite average consumption, the measurement can still proceed with certainty but will suffer a bounded error on its statistics. 

The structure of this paper is as follows. In \secr{framework} we layout the framework we shall use in this study and describe the approach to nonlocal measurements taken with specific attention paid to the Bell measurement example. This is followed in \secr{teleportjoint} by a brief review of teleportation as an ingredient in instantaneous protocols and a outline of the pioneering work by Vaidman~\cite{Vaidman03}. The main component of this work, what we call {\em rotation chains}, is introduced in \secr{finite}. In this section the protocol for a single chain is described in detail and is shown to have a finite average entanglement consumption. In addition it is explained how these chains can be concatenated to implement arbitrarily complex nonlocal unitaries and the scaling of the average entanglement consumption with the number of chains is also found. The remainder of the paper then utilises these tools for several nonlocal measurement problems. Firstly, in \secr{stateverify} it is applied to state verification measurements starting with an arbitrary two-qubit state before generalising to an arbitrary finite-sized bipartite multi-qubit system where the scaling of entanglement consumption with the Schmidt rank of target state is obtained. Secondly, the state verification scheme is expanded in \secr{operator_measure} to enable the simultaneous verification of any set of orthogonal eigenstates constituting a full operator measurement. Again two-qubit observables are considered in detail, followed by a bipartite multi-qubit system where the scaling in entanglement consumption with the system size is determined. Finally in \secr{conclusion} we conclude and comment on open problems for future work. 

\section{Framework}
\label{framework}
Let us now describe in more detail the framework used within this study. We shall exclusively consider both the principal system and measuring probes as being composed of two distinguishable parts built up from spin-$\half$ particles (qubits) and each localized in different regions of space occupied by Alice and Bob which are spacelike separated. While this is not the most general scenario it has proven to be particularly well suited for investigating quantum measurements and non-locality~\cite{Bohm,Bell,Popescu}. The local regions themselves are assumed to be small enough to neglect causality restrictions within them, but large enough compared to the Compton wavelength to neglect relativistic effects such as pair creation. Relativistic causality then enters due to the scale of the distances between the two parts of the system and otherwise the formalism of non-relativistic quantum mechanics can be used. Within this setting we shall consider measurement schemes which are localizable\footnote{Following earlier work~\cite{Beckman01} the relevance of our results for quantum field theory should be understood as applying to the idealization that the external probe variables are ``heavy'' with rapidly decaying correlations, while the field variables are ``light''. In this situation the notion of localizability, which requires a strict separation between field and probe, is credible.} quantum operations. This means that they can be composed of arbitrary local operations between the local parts of the system (we assume all local operations are equally easy to apply) and entangled resources which were shared prior to the measurement, but do not utilise any classical communication. Localizable operations are manifestly causal, although curiously not all causal operations are themselves localizable~\cite{Beckman01}.

\begin{figure}[t]
\begin{center}
\includegraphics[width=10.8cm]{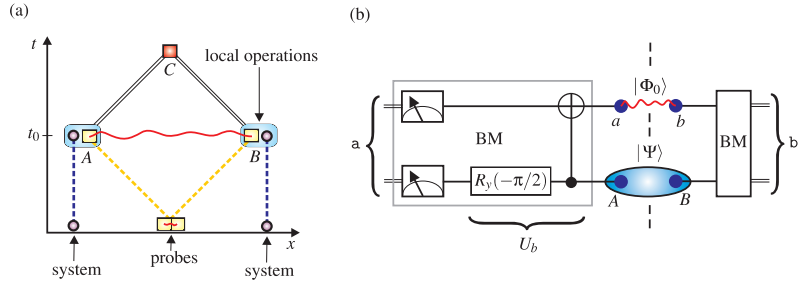}
\caption{(a) A depiction of the type of measurement scheme considered in this work. The scheme intends to measure a property, at a given time $t_0$, of a system composed of two spacelike separated parts at regions $A$ and $B$. To do so measuring probes are prepared some time earlier, possibly in an entangled state (signified by the wiggly line), and transported to the two locations $A$ and $B$. Once the probes arrive at these locations at time $t_0$ local operations are performed between the parts of the system and probe at each region and resulting in local classical information. This information is then transmitted to a location $C$ where the future lightcones of $A$ and $B$ intersect. The overall outcome of the instantaneous nonlocal measurement is then deduced at $C$, but pertains to the system at time $t_0$. (b) As an example of such a measurement scheme a circuit diagram is shown for the demolition nonlocal measurement of the Bell operator on two qubits utilizing one maximally entangled ancilla $\ket{\Phi_0}$. The vertical dashed line delineates the two regions and highlights that all operations in this circuit are local. In this example both local operations correspond to a Bell measurement. To aid the explanation of this measurement given in \secr{teleportjoint} we show on the lefthand side a Bell measurement composed of a unitary $U_b$ and single-qubit measurements~\cite{Nielsen} whose outcomes together give a binary encoding of the overall $\{0,1,2,3\}$ result. For subsequent diagrams we shall simply denote local Bell measurements by a ``BM'' box with four outcomes and not be concerned with its internals, be them a joint measurement projecting on to local Bell states or the single-qubit form given above. The {\em global} outcome for the instantaneous nonlocal Bell measurement is then designated by the addition modulo 4 of the local results ${\tt c} = {\tt a}\oplus{\tt b}$. }\label{frameworkfig}
\end{center}
\end{figure}

In general a nonlocal measurement requires previously arranged cooperative actions of Alice and Bob which can be broken into three steps. In the first step, which will be seen to be essential, suitably entangled ancilla systems must be prepared and distributed to the parties. Second, each party performs a local operation, such as unitaries and ideal (irreversible) projective measurements, on their part of the principal system and entangled ancillae. Third, the classical information extracted by both parties in the second step is transmitted to a central location $C$ where the readout of the result is completed\footnote{A more general scenario can permit quantum information to be transmitted. This would enable so called {\em exchange measurements}~\cite{Aharonov86} to occur where the principal system is swapped into the measuring device, essentially freezing its state, and is then later measured at $C$. In this case that the measurement has not really occurred until the last step and its outcome did not exist at time $t_0$.}. These steps are summarised in \fir{frameworkfig}(a).  Since the local operations which act on parts of the system and measuring device in step two can proceed without waiting or knowing the outcomes of actions performed by the other party they can in principle be performed in an arbitrarily small time. Thus when we speak of an ``instantaneous measurement'' we are referring to the particular Lorentz frame where both observers performed their actions at time $t_0$. Since we are interested in examining questions of causality, as opposed to covariance, we shall continue to use the terminology of quantum states and confine our description to this Lorentz frame. At the end of step two both Alice and Bob are in possession of a set of indelible local classical bits. In accordance with the information erasure theorem~\cite{Popescu} these local outcomes can only specify which eigenvalue of the nonlocal variable the system had at time $t_0$ once they are combined later at a point $C$ in the future light cones of both observers. As a consequence although the measurement was instantaneous and completed in step two at time $t_0$ the result is not necessarily known instantaneously by either party and can only be reconstructed much later at step three. Despite these features nonlocal verification measurements retain the usual requirements that (i) when the system is in an eigenstate of the observable the outcome corresponding to that eigenstate is produced with certainty, and the linearity of quantum mechanics then ensures that (ii) for a general superposition of eigenstates the corresponding eigenvalues are observed with the appropriate quantum probabilities.

The features of a nonlocal measurement just discussed are best outlined by a concrete example. In \fir{frameworkfig}(b) a nonlocal demolition measuring scheme for the Bell operator of two qubits is shown. The Bell operator possesses the non-degenerate maximally entangled eigenstates
\begin{eqnarray}
\ket{\Phi_0} = \frac{1}{\sqrt{2}}\left(\ket{0}_A\ket{0}_B + \ket{1}_A\ket{1}_B\right), \nonumber \\
\ket{\Phi_1} = \frac{1}{\sqrt{2}}\left(\ket{0}_A\ket{1}_B + \ket{1}_A\ket{0}_B\right) \nonumber \\
\ket{\Phi_2} = \frac{1}{\sqrt{2}}\left(\ket{0}_A\ket{0}_B - \ket{1}_A\ket{1}_B\right), \nonumber \\
\ket{\Phi_3} = \frac{i}{\sqrt{2}}\left(\ket{0}_A\ket{1}_B - \ket{1}_A\ket{0}_B\right). \nonumber
\end{eqnarray}
For the measurement scheme shown in \fir{frameworkfig}(b) a pair of ancilla qubits in the state $\ket{\Phi_0}$ have been previously distributed. Such maximally entangled pairs will form the resource for all of the schemes studied in this work. The measurement of the Bell operator then proceeds by each party performing a local Bell measurement between their half of the system and ancilla pair. Since it is a demolition verification measurement once it is completed the local parts of the system and ancilla are left in direct product states with equal unbiased probabilities. Thus in accordance with causality local information in the relevant degrees of freedom is erased and the local reduced density matrix is maximally mixed at all times. As a result the local outcomes ${\tt a}, {\tt b} \in \{0,1,2,3\}$ reveal no information about the global outcome in isolation. Instead they are correlated nonlocally with the final outcome being ${\tt c} = {\tt a}\oplus {\tt b}$ where $\oplus$ is modulo 4 addition. We shall explain how this measurement scheme works shortly in \secr{teleportjoint}. While this demolition Bell measurement scheme shares many features with the more general schemes about to be introduced we mention for completeness that, with the use of an additional entangled pair and a suitable modification of the circuit in \fir{frameworkfig}(b), a non-demolition scheme can be devised~\cite{Aharonov80,Aharonov81,Breuer}. The information erasure theorem~\cite{Popescu} proves that this is the only nonlocal variable of two qubits which possesses a non-demolition measurement scheme because the reduced density matrix of any of its eigenstates for either party is maximally mixed.

\section{Teleportation and instantaneous nonlocal unitaries}
\label{teleportjoint}
To generalise the Bell measurement just described to a more general nonlocal measurement scheme it turns out to be very convenient to describe the local operations performed by both parties in terms of the instantaneous part of the teleportation protocol~\cite{Bennett93}. In this section we shall describe teleportation within the framework outlined above and also detail earlier work by Vaidman~\cite{Vaidman03} which demonstrated how, through a prearranged recursive structure, it enables the instantaneous measurement of any nonlocal variable.

\subsection{Teleportation}
\label{teleportation}
As is well known the teleportation~\cite{Bennett93} of an arbitrary state of $d$ qubits $\ket{\Psi}$ can be accomplished by local operations and classical communication if Alice and Bob share one half of $d$ maximally entangled two-qubit states $\ket{\Phi_0}$. This follows from the identity
\begin{eqnarray}
\ket{\Psi}_{A_1A_2\cdots
A_d}\otimes\ket{\Phi_0}_{a_1b_1}\otimes\ket{\Phi_0}_{a_2b_2}\otimes\cdots\otimes\ket{\Phi_0}_{a_db_d}
 \nonumber \\
= \frac{1}{2^d}\sum_{\bf
m}\ket{\Phi_{{\tt a}_1}}_{A_1a_1}\otimes\ket{\Phi_{{\tt a}_2}}_{A_2a_2}\otimes\cdots\otimes\ket{\Phi_{{\tt a}_d}}_{A_da_d}
\sigma_{\bf a}\ket{\Psi}_{b_1b_2\cdots b_d}. \nonumber
\end{eqnarray}
Here we have designated a tensor product of Pauli operators\footnote{We will refer to tensor products of Pauli operators as a Pauli string operator or a simply as a Pauli distortion depending on the context.} over the system of $d$ qubits as $\sigma_{\bf a} = \sigma_{{\tt a}_1}\otimes\sigma_{{\tt a}_2}\otimes\cdots\otimes\sigma_{{\tt a}_d}$, where ${\bf a} = ({\tt a}_1,{\tt a}_2,\cdots,{\tt a}_d)$ is an $d$-dimensional vector of outcomes ${\tt a}_j \in \{0,1,2,3\}$ and we numerically index $\sigma_{\tt k}$ with $\sigma_0 = \mathbbm{1}$ while $ 1 \mapsto x, 2 \mapsto z$ and $3 \mapsto y$. Teleportation is then achieved by Alice measuring each pair of qubits $A_j,a_j$ in the Bell basis $\ket{\Phi_{\tt k}}$ with the outcome fixing the element ${\tt a}_j$ in $\bf a$. Overall this collapses Bob's $d$ qubits to the state $\ket{\Psi}$, modulo a Pauli distortion $\sigma_{\bf a}$ determined by Alice's equiprobable measurement outcomes $\bf a$. The full teleportation protocol is finished by Alice transmitting $2d$ classical bits to Bob specifying the vector of outcomes $\bf a$ so he can remove the Pauli distortion $\sigma_{\bf a}$  and recover $\ket{\Psi}$ with certainty. Since we will be exclusively concerned with instantaneous operations this last step, which necessarily takes a finite amount of time to implement, will never be performed and we shall from now on use the term {\em teleportation} to describe the Bell measurement part only. The cost of instantaneity is the unavoidable presence of the equiprobable Pauli distortion $\sigma_{\bf a}$ which due to the identity
\begin{eqnarray}
 \frac{1}{4^d}\sum_{\bf a} \sigma_{\bf a}\outprod{\Psi}{\Psi}\sigma_{\bf a} &=& \frac{1}{2^d}\mathbbm{1}, \nonumber
\end{eqnarray}
preserves relativistic causality by completely scrabbling the reduced density matrix of the local system. As we shall see despite the distortion teleportation can nonetheless be exploited to achieve instantaneous non-local operations.

\begin{figure}[t]
\begin{center}
\includegraphics[width=10cm]{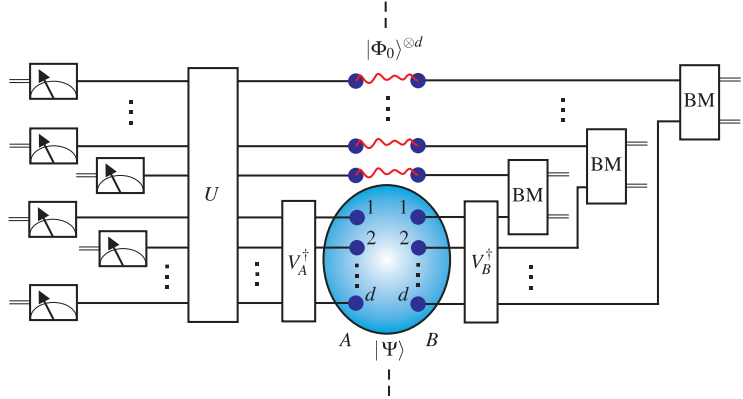}
\caption{A schematic of a nonlocal verification measurement scheme for any observable on $2d$ qubits whose complete set of eigenstates are locally equivalent to stabiliser states. The unitaries $V_A^\dagger$ and $V_B^\dagger$ are arbitrary and account for the local equivalence. The unitary $U \in \mathcal{S}$ is a stabiliser and is applied to the system once it has been localised by teleportation. The entire set of qubit at Alice's location can be measured in the $z$-axis to complete the scheme. Generalisation to unequal distributions of qubits and multi-party scenarios is straightforward.}\label{stabiliser_fig}
\end{center}
\end{figure}

The simple instantaneous Bell measurement in fact already highlights some essential properties of nonlocal measurements and readily allows us to identify a bipartite multi-qubit generalisation. Specifically in \fir{frameworkfig}(b) we can interpret Bob's local Bell measurement (on the right) as a teleportation of his half of the system to Alice yielding an outcome {\tt b}. In a step which will be shared by all schemes in this work he has localised their initially distributed system. Alice can then attempt to apply a unitary which maps the locally unmeasurable set of eigenstates into the trivially measurable direct product set $\ket{0}\ket{0}, \ket{0}\ket{1}, \ket{1}\ket{0}$ and $\ket{1}\ket{1}$. Were Alice to apply a general unitary $U$ its effect would be confounded by the Pauli distortion $\sigma_{\tt b}$ on her receiving qubit unknown to her. For a Bell measurement the required unitary $U_b$, depicted in \fir{frameworkfig}(b), is a member of a special class of unitaries in this regard. Specifically, for a multi-qubit system with a distortion $\sigma_{\bf b}$ there are a set of unitaries $U \in \mathcal{S}$ which satisfy $U\sigma_{\bf b} = \sigma_{{\bf b}'}U$, in which a Pauli distortion $\sigma_{\bf b}$ can be propagated through them at the expense of possibly changing to a different Pauli distortion $\sigma_{{\bf b}'}$. This set of unitaries $\mathcal{S}$ are called stabilisers and can be constructed, up to a global phase, from quantum circuits containing only CNOT, Hadamard and phase gates~\cite{Nielsen}.

Since both the CNOT gate and the rotation $R_y(-\pi/2) = \exp(i\pi\sigma_y/4)$ are stabilisers the effect of $U_b(\mathbbm{1}\otimes\sigma_{\tt b})$ in the Bell measurement is summarised as $U_b$, $(\mathbbm{1}\otimes\sigma_x) U_b$, $(\sigma_x\otimes\sigma_z) U_b$, and $(\sigma_x\otimes\sigma_y) U_b$ for ${\tt b} = (0,1,2,3)$, respectively. The scheme terminates, once $U_b$ is applied, with a measurement of both qubits in the computational basis (i.e. $z$-axis). Since Pauli distortions simply map direct product states between themselves once $\sigma_{\tt b}$ has been propagated through $U_b$ it induces a benign, but causality preserving, non-deterministic mapping between the Bell and direct product bases. This final measurement in the fixed $z$-axis is therefore certain to complete the scheme.  

With this observation we can immediately construct a nonlocal instantaneous verification measurement for any operators on any number of qubits whose eigenstates are all stabiliser states (or states which are locally equivalent to them). In addition to Bell states this class includes some of the most well studied multi-qubit entangled states such as the Greenberger-Horne-Zeilinger state~\cite{Greenberger}, cluster states~\cite{Raussendorf01,Raussendorf03} and more generally graph states~\cite{Hein}. The entanglement consumption of stabiliser measurements, analogous to the Bell measurement, is then simply the minimum number of ebits needed to localise the system. A schematic diagram of a stabiliser measurement protocol is given in \fir{stabiliser_fig}. For the most general nonlocal measurements the unitary $U$ required will not be a stabiliser. Our goal is therefore to devise a scheme which, after the system has been localised by teleportation, enables $U$ to be applied while still propagating any Pauli distortions to the end. As we shall see for arbitrary unitaries $U$ this is a highly nontrivial and expensive task.

\subsection{The Vaidman scheme}
\label{vaidmanscheme}
The general nonlocal measurement scheme devised by Vaidman~\cite{Vaidman03} starts in the same way as Bell measurement in \fir{frameworkfig}(b) by Bob teleporting his half of the system to Alice. Without any loss of generality we focus  on a system of two qubits. Since Alice and Bob's aim is to measure some nonlocal variable $O$ with eigenstates $\ket{o_1}, \ket{o_2}, \ket{o_3}$ and $\ket{o_4}$ they devise a unitary $U$ transformation which maps these eigenstates to the measurable direct product basis as
\begin{eqnarray}
U\ket{o_1} &=& \ket{0}\ket{0}, \quad U\ket{o_2} = \ket{0}\ket{1}, \nonumber \\
U\ket{o_3} &=& \ket{1}\ket{0}, \quad U\ket{o_4} = \ket{1}\ket{1} \nonumber.
\end{eqnarray}
Given the system was initially in that state $\ket{\Psi}_{AB}$ the state of Alice's qubit $A$ and the ancilla qubit $a_1$, representing the receiving qubit of the teleportation from Bob, is now in a state $\sigma_{{\tt b}_1}\ket{\Psi}_{Aa_1}$. The first step of the scheme is for Alice to simply apply $U$ to qubits $A$ and $a_1$. With a probability of $1/4$ Bob's teleportation will be non-distorting with ${\tt b}_1=0$ and Alice would have successfully mapped the eigenstates of $O$ to the measurable direct product basis. For the other three distortions the resulting unitaries $U\sigma_x$, $U\sigma_y$ and $U\sigma_z$ will not in general map the eigenstates to the direct product basis (unless of course $U$ happens to be a stabiliser). Since Alice has no knowledge of ${\tt b}_1$ she has no choice but to teleport the entire system of two qubits back to Bob. For brevity we shall from now on call a complete teleportation of the system, regardless of the number qubits, a {\em channel}. This initial step of the scheme is shown in \fir{vaidman_fig}(a).

\begin{figure}[t]
\begin{center}
\includegraphics[width=13.68cm]{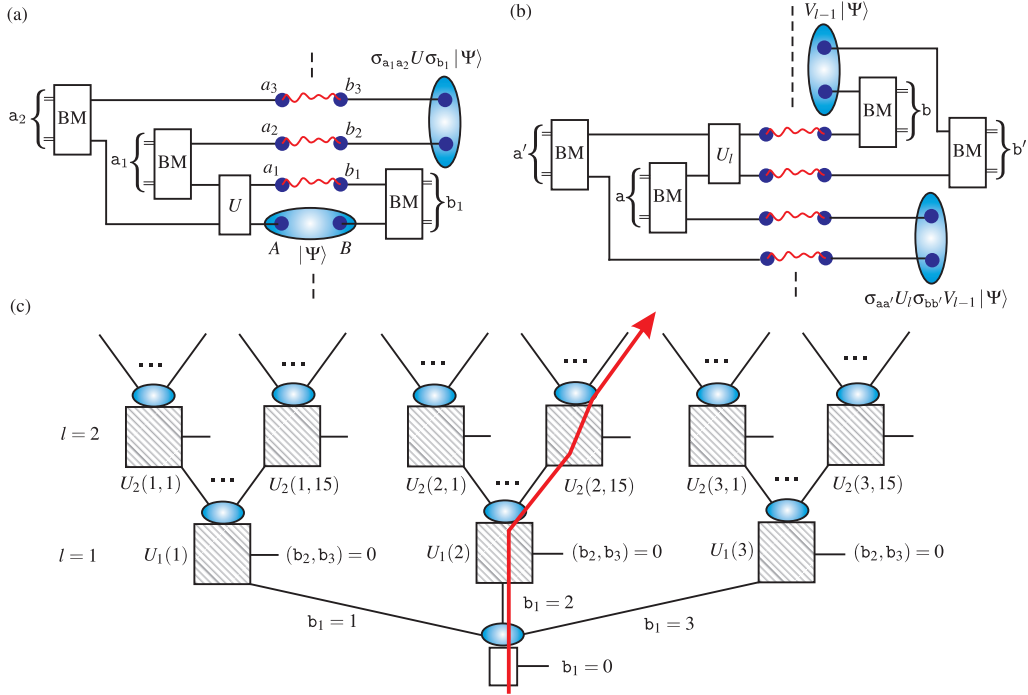}
\caption{(a) At the start of the Vaidman scheme Bob teleports his half of the system to Alice who then applies the unitary $U$ between her half and the teleported qubit. Since Alice does not know whether this action was successful she teleports the entire system back to Bob. (b) A cluster mentioned in the main text. Given Bob has a state $V_{l-1}\ket{\Psi}$ with some accumulative unitary $V_{l-1}$ applied to it he can teleport the entire system to Alice who can apply a correction $U_l$ and teleport it back. If Bob's teleportation is non-distorting so ${\tt b}={\tt b}'=0$ then Alice has successfully corrected the returned state. (c) A tree diagram of the Vaidman scheme. The root of the tree is the initial step of the scheme depicted in (a). The hashed boxes spawning from the root represent clusters identical to that depicted in (b). Depending on his teleportation outcomes Bob traverses the tree structure utilising only specific path of clusters corresponding to his history of outcomes (e.g. illustrated by the arrow). See the main text for a more detailed description of the scheme.}\label{vaidman_fig}
\end{center}
\end{figure}

At Bob's side he expects the return of the system and on the fortuitous occasion that his first teleportation gave ${\tt b}_1=0$ he can be assured that Alice successfully applied $U$ to $\ket{\Psi}$ leaving his two ancilla qubits $b_2$ and $b_3$ in the state $\sigma_{{\tt a}_1{\tt a}_2}U\ket{\Psi}_{b_2b_3}$. Just as with the Bell measurement scheme the final mapping is to the trivial direct product basis modulo a subsequent Pauli distortion. Thus, despite the fact that Bob has no knowledge of the outcomes ${\tt a}_1$ and ${\tt a}_2$ he can go ahead and immediately measure the qubits in the $z$-axis completing the verification measurement of $O$. For the cases where ${\tt b}_1 \neq 0$ Bob knows that Alice did not apply $U$ in isolation. To allow Alice the opportunity to correct this mistake the scheme from here on adopts a tree-like structure. The root of this tree is the first teleportation just described. Above this there are now three branches each labelled by the possible distorting values ${\tt b}_1$ might take. Each branch leads to a {\em cluster} whose structure is illustrated in \fir{vaidman_fig}(b). A cluster simply contains a teleportation channel for Bob to send back the system to Alice and a corresponding channel for Alice to return it. Depending on the actual value of ${\tt b}_1$ Bob traverses the corresponding branch of the tree and sends the system back to Alice via the channel in that branches cluster. He will never act on clusters in any other branches of the scheme.

At the receiving end of the incoming channel in each of the three clusters Alice will have, if the cluster was used, a state $\sigma_{{\tt b}_2{\tt b}_3}\sigma_{{\tt a}_1{\tt a}_2}U\sigma_{{\tt b}_1}\ket{\Psi}$. She can now infer the value of ${\tt b}_1$ from the cluster's label. Under the assumption that Bob's teleportation in that cluster was non-distorting, so ${\tt b}_2={\tt b}_3=0$, she can devise a correction unitary $U_1({\tt b}_1)$, dependent on ${\tt b}_1$, obeying
\begin{eqnarray}
U_1({\tt b}_1)\sigma_{{\tt a}_1{\tt a}_2}U\sigma_{{\tt b}_1} &=& U. \nonumber 
\end{eqnarray}
Thus, with a probability of $1/16$ Alice will undo the previous unitary and distortions and map the eigenstates of $O$ to the direct product basis. To complete the cluster she teleports the resulting two qubit system back to Bob via a corresponding return channel as shown in \fir{vaidman_fig}(b). Since Alice does not know which, if any, of the clusters were used she must perform this ${\tt b}_1$ dependent correction on all three clusters. 

The situation for Bob is now identical to the first round but with a smaller probability of success. If ${\tt b}_2={\tt b}_3=0$ then as before he can immediately measure the incoming qubits on the cluster he used and complete the measurement. For the other fifteen possible distorting outcomes Bob knows Alice's correction will have failed. To overcome this failure the same strategy is applied. Each of the clusters in the first level of the tree spawn fifteen new branches, one for each possible distortion by Bob's previous teleportation, again leading to a new cluster. From his current position in the tree Bob now traverses the appropriate branch dependent on ${\tt b}_2,{\tt b}_3$ and teleports the system back to Alice through the channel in that branch's cluster. For Alice the situation is now that she has 45 incoming channels to operate on since she has no knowledge of Bob's actual path through the tree. For each cluster in this second level she can continue to guarantee a 1/16 chance of success by again devising a unitary $U_2({\tt b}_1,{\tt b}_2,{\tt b}_3)$ obeying
\begin{eqnarray}
U_2({\tt b}_1,{\tt b}_2,{\tt b}_3)\sigma_{{\tt a}_4{\tt a}_5}U_1({\tt b}_1)\sigma_{{\tt b}_2{\tt b}_3}\sigma_{{\tt a}_1{\tt a}_2}U\sigma_{{\tt b}_1} &=& U. \nonumber 
\end{eqnarray}
The labels on the tree structure provide Alice with a complete history of distortions which Bob would have induced had he traversed those branches and this is essential for her to be able to construct a correction. The only knowledge Alice lacks is the nature of Bob's last teleportation and her correction only works on the assumption that it is non-distorting. The scheme therefore continues in the same way following an exponentially growing tree structure, depicted in \fir{vaidman_fig}(c). The measurement is completed once Bob has performed a non-distorting teleportation. 

So long as this scheme is repeated to infinite depth it can, quite remarkably, ensure that at some point along the path traversed by Bob the unitary $U$ is applied with certainty, modulo some proceeding Pauli distortions. At this termination point Bob can then complete the measurement. The cost of achieving this task though is unbounded. In particular the division of labour is highly skewed since Alice must operate on all branches, whose number grows exponentially with the level, and unlike Bob has no termination condition. This means that  the infinite amount of entanglement which was initially distributed to form the scheme's tree structure is necessarily consumed without exception. In return for this effort, however, Alice has complete control of the unitary $U$ eventually implemented and need only decide what it is immediately before she starts her actions. The scheme generalises straightforwardly for $d$ qubits but with a probability of success $4^{-d}$ at each level and $4^d-1$ new branches spawning to the next level. Additionally since Alice does all the correction work the scheme can be readily adapted to work with any number of parties~\cite{Vaidman03}. This is a property shared with stabiliser measurements since there only one party is needed to perform the stabiliser circuit. The Vaidman scheme provides a constructive proof that an instantaneous measurement scheme, which is guaranteed to succeed, can be devised for any nonlocal variable and nonetheless be compatible with both quantum mechanics and causality. For the remainder of this work we describe a scheme, based on a simple but significant modification of Vaidman's, that can similarly be used to measure any nonlocal variable and guaranteed to succeed, but consumes only a finite amount of entanglement on average.

\section{Finite consumption scheme}
\label{finite}
The essential adjustment we make to Vaidman's scheme is that rather than attempting to apply the desired unitary $U$ directly at each step we instead decompose $U$ into a sequence of simpler unitaries and attempt to apply these individually in separate rounds of the scheme. These simpler unitaries are Pauli rotations $R_{\bf j}(\theta) = \exp(-i \theta \sigma_{\bf j}/2) = \cos(\half\theta)\mathbbm{1} - i\sin(\half\theta)\sigma_{\bf j}$ involving the exponential of a Pauli string operator $\sigma_{\bf j}$, introduced earlier in \secr{teleportation}, by an angle $\theta$. In this section we will concentrate on implementing Pauli rotations and give explicit examples of decomposing general unitaries $U$ in terms of them later when we discuss specific applications in \secr{stateverify} and \secr{operator_measure}. Despite Pauli rotations not being stabilisers (aside from when $\theta = \pi/2$) they do have extremely advantageous properties with respect to Pauli distortions. As we shall now describe this can be exploited to yield a scheme where both parties have local termination conditions and only a finite amount of the initial entanglement is ever consumed.

\subsection{Pauli rotation chain}
\label{rotation_chain} 
The basic component of all our measurement schemes is a {\em rotation chain} which applies a single Pauli rotation $R_{\bf j}(\theta)$ designated by an angle $\theta$ and a nontrivial vector $\bf j$ specifying the Pauli string known to both parties.  A rotation chain is composed of a sequence of teleportation channels in which the entire system of $d$ qubits is teleported together back and forth in an alternating direction between Alice and Bob, as depicted in \fir{rotation_chain_fig}. The starting point of a rotation chain is the familiar situation where one party, say Alice, possesses the entire system. Initially, when the system was distributed, it was in some state $\ket{\Psi}$, however, the actual state of the system at Alice's location contains a Pauli distortion $\sigma_{{\bf b}_1}\ket{\psi}$ defined by a vector ${\bf b}_1$ known only to Bob. This distortion is taken to have arisen from earlier teleportations (such as previous rotation chains as described shortly in \secr{concatenate}) and so we take all $4^d$ possible vectors  ${\bf b}_1$ as equiprobable\footnote{For Bob's initial teleportation which localises the system $\sigma_{{\bf b}_1}$ has zero elements for all Alice's qubits and so is only equiprobable over a subset of $4^{d/2}$ strings. However, since Alice will attempt to apply a Pauli rotation on the entire system the effect of this type of $\sigma_{{\bf b}_1}$ is identical.}. As we have seen the presence of this distortion generally results in $4^d-1$ different errors if Alice tried to apply the complete unitary $U$ directly. In a rotation chain Alice instead applies $R_{\bf j}(\theta)$ and only one of two possibilities occur
\begin{eqnarray}
R_{\bf j}(\theta)\sigma_{{\bf b}_1}\ket{\Psi} &=& \left\{
\begin{array}{r@{\,}l} \sigma_{{\bf b}_1} R_{\bf
j}(\theta)\ket{\Psi}, \quad{\rm for}\quad & {\bf b}_1 \in {\bf
c}({\bf j})
\\ \sigma_{{\bf b}_1} R_{\bf
j}(-\theta)\ket{\Psi}, \quad{\rm for}\quad &  {\bf b}_1 \in
\bar{{\bf c}}({\bf j}) \end{array}. \right. \label{one_qubit_op}
\end{eqnarray}
Since ${\bf j} \neq (0,0,\cdots,0)$, and so never designates a string of identity operators, we denote here ${\bf c}({\bf j})$ as the set of $4^{d}/2$ vectors specifying Pauli strings which commute with $\sigma_{\bf j}$, while $\bar{{\bf c}}({\bf j})$ is the other half of the total set of vectors which anti-commute with $\sigma_{\bf j}$. In the latter case propagation of the rotation through the distortion $\sigma_{{\bf b}_1}$ results in a sign change. Thus, Alice has a probability of $\half$, independent on $d$, to have implemented the correct rotation on the initial state. Moreover the only error she can make is to rotate in the wrong direction. Since she has no knowledge of her success she must teleport the entire system back to Bob via the first channel shared between them.

\begin{figure}[t]
\begin{center}
\includegraphics[width=9.72cm]{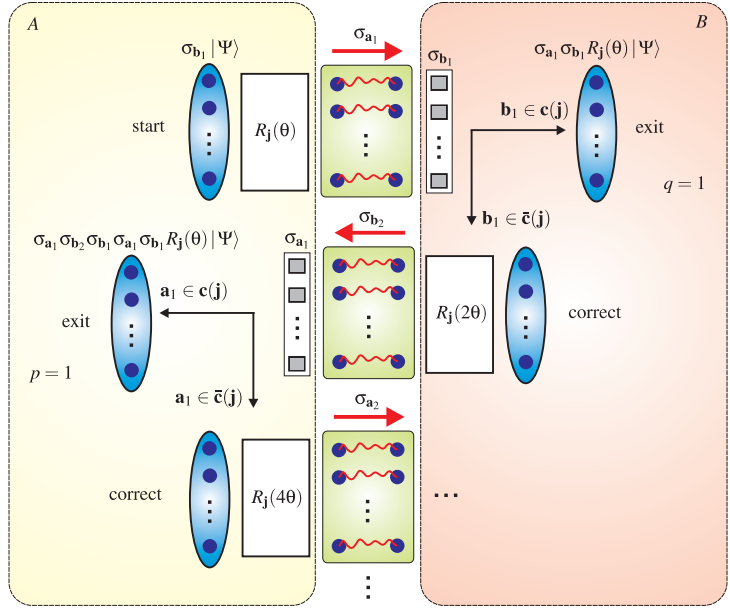}
\caption{A schematic of a Pauli rotation chain used to implement a unitary $R_{\bf j}(\theta)$. As a result of previous teleportations Alice possesses the entire system, but in a state $\sigma_{{\bf b}_1}\ket{\Psi}$, where ${\bf b}_1$ is known only to Bob. As a result she cannot be certain that she has applied $R_{\bf j}(\theta)$ directly to the state $\ket{\Psi}$. The scheme depicted shows that by exploiting a sequence of directed teleportation channels the entire system can be bounced back and forth between Alice and Bob such that there is a probability of $\half$ at each step that either party possesses a state $R_{\bf j}(\theta)\ket{\Psi}$ modulo a proceeding Pauli distortion. This strategy is a bipartite multi-qubit generalization of a similar single-qubit scheme presented in~\cite{Vaidman03}. See the main text for a more detailed description of the scheme.}\label{rotation_chain_fig}
\end{center}
\end{figure}

At Bob's side he immediately applies the unitary $\sigma_{{\bf b}_1}$, corresponding to the initial distortion, to the incoming qubits. If ${\bf b}_1 \in {\bf c}({\bf j})$ then his initial distortion was commuting and the incoming qubits will be in a state $\sigma_{{\bf b}_1} \sigma_{{\bf a}_1} \sigma_{{\bf b}_1} R_{\bf j}(\theta)\ket{\Psi}$, where $\sigma_{{\bf a}_1}$ is a new distortion induced by Alice's teleportation. Bob therefore knows that the incoming qubits have had, modulo a subsequent distortion, the correct rotation applied to their initial state. He then keeps these qubits ready for further operations (see \secr{concatenate}) or a measurement. His actions for this chain are then terminated. If ${\bf b}_1 \in \bar{{\bf c}}({\bf j})$ then his initial distortion was anti-commuting and he knows that Alice performed $R_{\bf j}(-\theta)$ instead. Following a strategy outlined in~\cite{Cirac} Bob can attempt to correct this, under a previously agreed assumption that ${\bf a}_1$ is commuting, by applying a new {\em double angle} rotation $R_{\bf j}(2\theta)$ to the qubits. This gives $R_{\bf j}(2\theta)\sigma_{{\bf b}_1}\sigma_{{\bf a}_1}\sigma_{{\bf b}_1} R_{\bf j}(-\theta)\ket{\Psi}$ which is the desired state only if ${\bf a}_1 \in {\bf c}({\bf j})$ is a commuting distortion. To overcome his lack of knowledge regarding ${\bf a}_1$ Bob teleports all $d$ qubit back to Alice via the next channel in the chain.

The situation for Alice is now identical to Bob's just described. She immediately applies $\sigma_{{\bf a}_1}$ to the incoming qubits. If ${\bf a}_1 \in {\bf c}({\bf j})$ she can be certain that, if it was necessary, Bob succeeded in correcting her rotation. In this case the state of her system is $\sigma_{{\bf a}_1}\sigma_{{\bf b}_2}\sigma_{{\bf
b}_1}\sigma_{{\bf a}_1}\sigma_{{\bf b}_1} R_{\bf j}(\theta)\ket{\Psi}$, where $\sigma_{{\bf b}_2}$ is a new distortion induced by Bob's teleportation back. This final state is of the required form so she keeps the qubits and terminates her actions in this chain. If ${\bf a}_1 \in \bar{{\bf c}}({\bf j})$ Bob's rotation causes an accumulative error of $R_{\bf j}(-3\theta)$. Alice attempts to correct this, again under the assumption that his last distortion $\sigma_{{\bf b}_2}$ is commuting, by applying another double angle rotation $R_{\bf j}(4\theta)$. Notice that she does not need to assume or know anything about earlier distortions by Bob, such as $\sigma_{{\bf b}_1}$, since it appears twice in the accumulative distortion. She then teleports the qubits back via the next channel and the scheme continues. A schematic of these steps in the rotation chain scheme are given in \fir{rotation_chain_fig}.

Notice that both Alice and Bob have a probability of $\half$ of implementing the jointly agreed rotation at each step and can both determine their success by local outcomes. Since the actions of both parties terminate there is a zero probability that the chain continues indefinitely and so only a finite amount of the initial entanglement is ever consumed. A disadvantage of joint termination is that as a rotation chain proceeds both parties lose knowledge of where the appropriately transformed qubits finally reside. Instead the actual pathway taken by the system is only reconstructed by the combination of Alice and Bob's local classical records. The manner in which the rotation chain deals with the Pauli distortions caused by teleportation is very reminiscent of one-way quantum computing~\cite{Raussendorf01,Raussendorf03}. There the indeterminism of single qubit measurements used to drive the computation produces Pauli distortions at intermediate stages which, via minor adjustments in the subsequent operations, are propagated to the end of the computation. Their effect is then to simply alter the interpretation of the final output measurements. If further rotations are required then, as we shall show in the next section, distortions can continue to be propagated to the end.

\subsection{Concatenation of rotation chains}
\label{concatenate}
Let us now suppose Alice and Bob wish to apply a further rotation $R_{\bf k}(\xi)$ to the $d$-qubit state $R_{\bf j}(\theta)\ket{\Psi}$. To do this they can use a second rotation chain which applies  $R_{\bf k}(\xi)$ to the output from the first $R_{\bf j}(\theta)$. However, since the first rotation chain has multiple opportunities of terminating successfully on both Alice and Bob's side a second chain must be available separately for each of these exit points to cover all eventualities. This gives a tree structure of concatenated chains like that shown in \fir{rotation_concat_fig}. Following this figure suppose that Alice exits the first chain first on her $q$th opportunity. The $d$-qubits she then possesses will be in a state carrying a large accumulative distortion dependent on its history up to that point through the first chain as
\begin{eqnarray}
\sigma_{{\bf a}_q}\sigma_{{\bf b}_{q+1}}\sigma_{{\bf b}_{q}}\sigma_{{\bf a}_q}\sigma_{{\bf a}_{q-1}} \cdots
\sigma_{{\bf b}_2}\sigma_{{\bf b}_1}\sigma_{{\bf a}_1}\sigma_{{\bf
b}_1} R_{\bf j}(\theta)\ket{\Psi}. \label{accum_dist_state}
\end{eqnarray}
She can go ahead and engage these qubits with the designated second rotation chain for this exit point which, in an identical way to the first, will apply $R_{\bf k}(\xi)$. Since the accumulative distortion in \eqr{accum_dist_state} contains two of every previous distortion, except for $\sigma_{{\bf b}_{q+1}}$, the criterion for Alice's success in applying the second rotation $R_{\bf k}(\xi)$ is based only on Bob's last teleportation, via ${\bf b}_{q+1} \in {\bf c}({\bf k})$, and not on the complete history. This is in stark contrast to the Vaidman scheme.

\begin{figure}[t]
\begin{center}
\includegraphics[width=7cm]{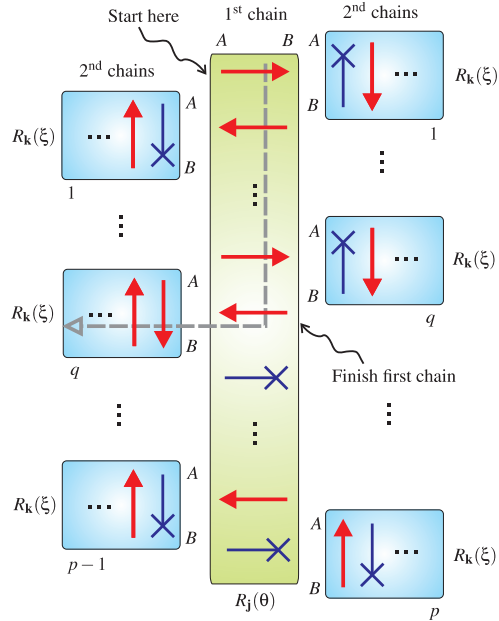}
\caption{A schematic of the concatenation of two Pauli rotation chains used to implement a unitary $R_{\bf k}(\xi) R_{\bf j}(\theta)$ modulo a proceeding Pauli distortion. For each possible exit from the $R_{\bf j}(\theta)$ rotation chain there is a second $R_{\bf k}(\xi)$ chain. The boxes containing arrows in this figure represent the entire rotation chain protocol depicted in \fir{rotation_chain_fig}. Arrows which end with a $\times$ indicate that the originating party has not participated in the protocol for this specific chain. While both parties participate in the first chain only one of the secondary chains has overlapping actions of Alice and Bob. In this figure Alice exits the first chain on her $q$th opportunity, while Bob exits on his $p$th where $p>q$. The dashed ``L''-shaped line indicates the actual path taken by the principal system in this case. Actions performed by either party not intersecting this line do not contribute to the final outcome, but the no-signalling restriction requires that they are performed so that all eventualities are covered and the desired unitary is implemented with certainty. }\label{rotation_concat_fig}
\end{center}
\end{figure}

As depicted in \fir{rotation_concat_fig} both Alice and Bob must perform all the necessary steps for each of the second chains covering all possible exit points of the other party up to the point where they themselves exit from the first chain. This ensures that if the other party was successful before them the overall scheme still succeeds with certainty. Since all the first chain and all those spawning from it have a zero probability of continuing indefinitely the overall scheme also has a finite average consumption. It is also clear that this concatenation can continue, albeit at increasing expense, for any finite sequence of rotations to be applied to the $d$-qubit initial state, and still retain a finite average entanglement consumption. We now examine more precisely what this consumption is.

\subsection{Average entanglement consumption}
\label{consumption}
To measure the consumption we count the number channels that are required on average. A detailed description of this calculation is given in \ref{consumption_calc}. In summary we find that the average channel consumption for a single rotation chain is $\av{c_1} = 5$, while concatenation of further rotation chains results in a rapid growth as $\av{c_2} = 20$, $\av{c_3} = 59$, $\av{c_4} = 156$ and so on. These channel averages $\av{c_n}$ give the average consumption of entanglement, measured in ebits, once they are multiplied by $d$. By utilising the recursive structure of the protocol the average channel consumption $\av{c_n}$ can be approximated, in the limit of a large number of concatenated rotations $n$, by the exponential growth
\begin{eqnarray}
\av{c_{n}} &\approx& C \phi^n \label{growth},
\end{eqnarray}
where $C = (10 + 7\sqrt{2})/4$ and $\phi = 1 + \sqrt{2}$. As shown in \fir{consumption_fig}(b) the fit of this approximation to the exact consumption for $n>4$ demonstrates that it is very good for all but the smallest $n$. 

We saw earlier that a rotation by an angle $\pi/2$ has the special property that $R_{\bf j}(\pi/2)$ is a stabiliser. In this case no rotation chain steps are required. More generally a chain involving a binary angle $\theta = \pi/2^D$ not only implements the desired rotation when a commuting distortion occurs but also when a sequence of $D-1$ erroneous rotations are made since the required double angle correction reduces to $R_{\bf j}(\pi/2)$. Thus for rotations with a binary angle the chain terminates with certainty in a finite number of teleportations~\cite{Groisman, Vaidman03}. An example of a single rotation chain applicable to an angle which is any odd multiple of $\pi/32$ is given in \fir{consumption_fig}(a). The total amount of initial channels which must be available for $n$ concatenated rotation chains, each of length $D$, is finite but grows exponentially with $n$ as
\begin{eqnarray}
C_{\textrm{init}} &=& \frac{(D-1)[(D-1)^n-1]}{D-2} \label{init_growth}.
\end{eqnarray}
In an identical way to the $D\rightarrow\infty$ case the average consumption $\av{c_n}$ of this initial resource can be computed. In \fir{consumption_fig}(b) both $C_\textrm{init}$ and $\av{c_n}$ are shown for $D=3$ and $D=7$. For $D\leq 3$ both the initial resource and the average consumption remain below the average consumption for $D\rightarrow\infty$. For $D>3$ the average consumption  $\av{c_n}$ rapidly converges to the $D\rightarrow\infty$ limit and the initial resources grow far beyond it.

While we have shown a finite average consumption in general a practically relevant question arises as to what effect the restriction to finite initial resources has for general rotations. One strategy for doing this is to simply truncate continuous angle rotation chains to some maximum number of iterations. Indeed if this strategy is applied to the Vaidman scheme, by limiting its tree-depth, it results in it having a finite consumption equal to its finite initial resources. This approach, however, introduces a possibility that the measurement will fail completely and yield no result. Binary angle rotation chains present a more elegant means of exploring the implications of finite initial resources for our scheme. Rather than truncating continuous angle rotation chains, we instead consider a more interesting and relevant scenario where the desired rotation angle is discretised to a multiple of a binary angle that matches the maximum allowed number of iterations. Given the decomposition of the desired final unitary $U$ into a sequence of Pauli rotations the new scheme performs rotations about the nearest binary angle $\tilde{\theta} = \lfloor2^D\theta/\pi\rceil \pi/2^D$ to the exact angle $\theta$. In this way we obtain a measurement scheme constructed from finite initial resources,  consuming only a fraction of those on average, and is guaranteed to succeed at the expense of only implementing an approximation of $U$. Since the finite scheme no longer maps the eigenstates of our desired observable $O$ to the direct product basis a crucial question is then how much the measurement statistics of our approximation differ from the exact case. For a single rotation $R_{\bf j}(\theta)$ the error can be defined as
\begin{eqnarray}
E(\theta,\tilde{\theta}) &=& \max_{\ket{\Psi}} \left \| \left(R_{\bf j}(\theta) - R_{\bf j}(\tilde{\theta})\right)\ket{\Psi} \right \|, \nonumber \\
&=& \max_{\ket{\Psi}} \left \| \left(\mathbbm{1} - e^{-\frac{i}{2}\Delta\theta\sigma_{\bf j}}\right)\ket{\Psi} \right \| \nonumber,
\end{eqnarray}
where the maximum is taken over all normalised states $\ket{\Psi}$ and $\Delta\theta = \tilde{\theta} - \theta$. The error $E(\theta,\tilde{\theta})$ can be shown~\cite{Nielsen} to bound the absolute difference between the probabilities $P$ and $\tilde{P}$ for the outcome of any positive operator valued measurement on $R_{\bf j}(\theta)\ket{\Psi}$ and $R_{\bf j}(\tilde{\theta})\ket{\Psi}$, respectively, as $|P - \tilde{P}| \leq 2 E(\theta,\tilde{\theta})$. An upper bound to $E(\theta,\tilde{\theta})$ can be obtained by assuming the maximum deviation for $\Delta\theta = \pi/2^{D+1}$ which gives
\begin{eqnarray}
E(\theta,\tilde{\theta}) &\leq& \sqrt{2}\sqrt{1 - \cos\left(\frac{\pi}{2^{D+2}}\right)} \approx \frac{\pi}{4}2^{-D} \nonumber
\end{eqnarray}
and shows that the error decreases exponentially with $D$. For a sequence of $n$ rotation chains implementing the binary approximation to $U$ an important result from quantum computation~\cite{Bernstein} shows that the overall error is at most the sum of the errors of the individual rotations and so the exponential suppression of the measurement error is retained. We will now finish this work by applying rotation chains to a variety of basic measurement problems. Our results will mostly concentrate on the average entanglement consumption of continuous angle rotation chains but can be equally viewed as an upper bound to the average consumption of any finite binary angle scheme.

\begin{figure}[t]
\begin{center}
\includegraphics[width=10.8cm]{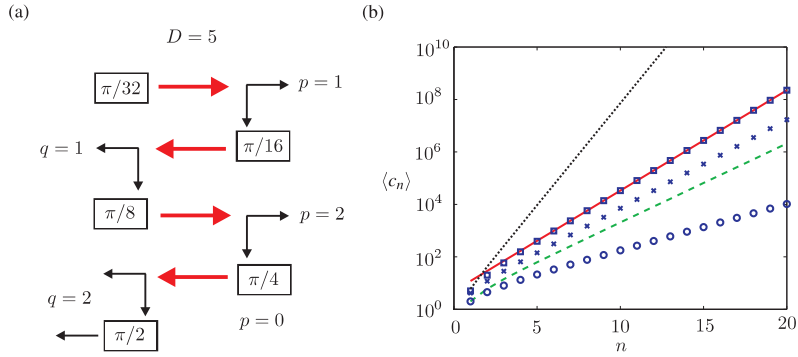}
\caption{(a) An example of a rotation chain for a binary angle $\theta = \pi/2^D$ with $D=5$. The scheme is guaranteed to terminate on Alice's second step $q=2$ since the correction is a rotation by $\pi/2$ which always succeeds modulo a proceeding Pauli distortion. For $D$ odd there is an outcome $p=0$ corresponding to when Bob never succeeds. (b) The average channel consumption $\av{c_n}$ for the scheme performing $n$ successive rotations of the form $R_{{\bf j}_n}(\theta_n)\cdots R_{{\bf j}_1}(\theta_1)$. The exact calculation of $\av{c_n}$ for non-binary $\theta_j$ angles (i.e. infinite length rotation chains) is shown ($\square$) as well as the pure exponential approximation (solid line) given in \eqr{growth}. The exact $\av{c_n}$ is also shown for binary angles with $D=3$ ($\circ$) and $D=7$ ($\times$), along with the total amount of initial channels $C_{\textrm{init}}$ which must be available in both cases as the (dashed line) and (dotted line), respectively.}\label{consumption_fig}
\end{center}
\end{figure}

\section{State verification measurements}
\label{stateverify}
Our first application of the tools developed in \secr{finite} is to state verification measurements. A verification of a given state $\ket{\Psi}$ means that the measurement always yields a ``yes'' result if the system is in the state $\ket{\Psi}$ and a ``no'' result if the system is in any orthogonal state $\ket{\Psi_\perp}$. If the initial state is a superposition then the appropriate probabilities for ``yes'' and ``no'' results follow from the linearity of quantum mechanics.  No assumptions are made about the final state of the system so there is no requirement that $\ket{\Psi}$ itself is undisturbed by the verification measurement. 

\subsection{Two-qubits states}
\label{twoqubit_verify}
To begin we present a simple scheme which performs a demolition verification of any two-qubit state $\ket{\Psi} \in \mathbbm{C}^2\otimes\mathbbm{C}^2$ split between two parties $A$ and $B$. The construction of a verification scheme for $\ket{\Psi}$ follows from its corresponding Schmidt decomposition
\begin{eqnarray}
\ket{\Psi} &=& \cos\left(\half\theta\right)\ket{\phi_0}_A\ket{\phi_0}_B + \sin\left(\half\theta\right)\ket{\phi_1}_A\ket{\phi_1}_B, \label{two_qubit_state}
\end{eqnarray}
where $\kets{\phi_k}$ are Alice's (Bob's) local Schmidt states and we have parameterized the corresponding Schmidt coefficients according to an angle $0\leq\theta\leq\pi/2$. To verify $\ket{\Psi}$ our scheme implements the inverse of the quantum circuit, shown in  \fir{two_qubit_verify}(a), that prepares $\ket{\Psi}$ locally. Starting from the initial state $\ket{0}_A\ket{0}_B$ this circuit performs a rotation $R_y(\theta)$ into the state $\ket{\Lambda_1} = \cos(\half\theta)\ket{0} + \sin(\half\theta)\ket{1}$ for qubit $A$, applies a CNOT gate $U_{cn}$ between the pair of qubits controlled by $A$, and is then followed by the product of single-qubit unitaries $V_A\otimes V_B$ which map the computational basis of each qubit into the respective local Schmidt basis as $\ket{k} \mapsto \ket{\phi_k}$, with $k \in \{0,1\}$. 

To invert this the verification scheme therefore starts with Alice and Bob performing the local unitary transformations $V^\dagger_A$ and $V^\dagger_B$. Bob then teleports his half of the system $B$ to Alice leaving qubit $A$ and her ancilla qubit $a$ in her possession in the distorted state $\sigma_{0{\tt b}} V^\dagger_A\otimes V^\dagger_B\ket{\Psi}_{Aa}$ described by his Bell measurement outcome ${\tt b}$. This is shown in \fir{two_qubit_verify}(b) and is labelled as step (i). Alice now applies a CNOT gate between qubits $A$ and $a$. Since the CNOT gate is a stabiliser any distortion can be propagated past it at the expense of spreading the distortion over the control qubit. Regardless of this Alice can be certain that she has implemented, up to a distortion, $U_{cn}V^\dagger_A\otimes V^\dagger_B\ket{\Psi_0} = \ket{\Lambda_1}_A\ket{0}_a$ and disentangled qubit $A$ from qubit $a$. She can then measure qubit $a$ completing step (ii) in \fir{two_qubit_verify}(b). The distortion $\sigma_{0{\tt b}}$ ensures that the outcome reveals no information to Alice.

\begin{figure}[t]
\begin{center}
\includegraphics[width=10.8cm]{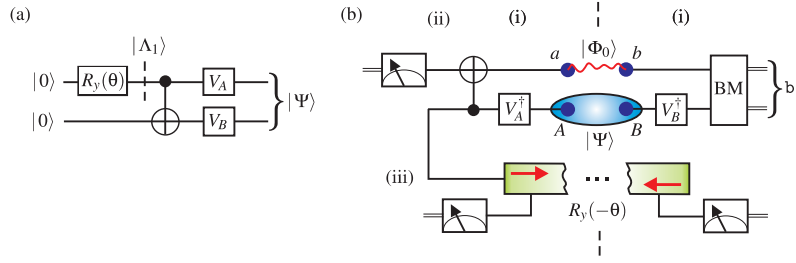}
\caption{(a) A local circuit which constructs an arbitrary two qubit state $\ket{\Psi}$ from a standard initial state $\ket{0}\ket{0}$. Firstly a rotation $R_y(\theta)$ is applied to qubit $A$ forming a single qubit state $\ket{\Lambda_1}$ composed of a superposition with real amplitudes corresponding to the Schmidt coefficients of $\ket{\Psi}$. This is then followed by a CNOT gate controlled by qubit $A$ and then two arbitrary single-qubit unitaries $V_A$ and $V_B$ are applied which rotate the computational basis into the required local Schmidt basis of $\ket{\Psi}$. (b) A nonlocal instantaneous verification of the state $\ket{\Psi}$ essentially reverses the circuit shown in (a). In step (i) the inverses of the local unitaries $V_A$ and $V_B$ are applied and the qubit $B$ is teleported to Alice. In step (ii) Alice then applies the CNOT and measures out the received qubit. The most complicated step is (iii) where the rotation $R_y(-\theta)$ is applied to qubit $A$. This is implemented via a single-qubit rotation chain followed by a measurement of the successful output.}\label{two_qubit_verify}
\end{center}
\end{figure}

Alice must now map the remaining qubit $A$, with certainty, into the $z$-axis so it too can be measured. To achieve this she needs to apply a rotation $R_y(-\theta)$. Her situation is identical to the scenario considered in \secr{rotation_chain} and can be readily dealt with using one single-qubit rotation chain as shown in step (iii) of \fir{two_qubit_verify}(b). The average entanglement $\av{e}$ consumed by this nonlocal two-qubit state verification scheme has no dependence on the value of $\theta$ except when it is a binary angle. In particular for a maximally entangled state with $\theta = \pi/2$ precisely 1 ebit is required, while the partially entangled states with $\theta = \pi/4$ or $\theta = \pi/8$ need precisely 2 and 3 ebits to be verified, respectively. Binary angles $\theta = (2m-1)\pi/2^D$, with $m$ integer, have a consumption
\begin{equation}
\av{e_D} = 6 + 2^{2-D} + 2^{-D/2}\left\{\frac{7}{\sqrt{2}}[-1 + (-1)^D] - 5[1 + (-1)^D]\right\}. \label{binaryconsump}
\end{equation} 
For any angle $\theta$ that is not binary $\av{e} = 6$ ebits on average and is independent of the entropy of entanglement of the state $\ket{\Psi}$. As expected this consumption is the asymptotic limit $D\rightarrow\infty$ of \eqr{binaryconsump}.

Although this measurement scheme was devised to verify a single state $\ket{\Psi}$ the ``no'' results do in fact verify a special set of states in the orthogonal complement,
\begin{eqnarray}
\ket{\Psi_1} = \cos\left(\half\theta\right)\ket{\phi_0}_A\ket{\phi_1}_B + \sin\left(\half\theta\right)\ket{\phi_1}_A\ket{\phi_0}_B, \nonumber \\
\ket{\Psi_2} = \sin\left(\half\theta\right)\ket{\phi_0}_A\ket{\phi_0}_B - \cos\left(\half\theta\right)\ket{\phi_1}_A\ket{\phi_1}_B, \nonumber \\
\ket{\Psi_3} = \sin\left(\half\theta\right)\ket{\phi_0}_A\ket{\phi_1}_B - \cos\left(\half\theta\right)\ket{\phi_1}_A\ket{\phi_0}_B,  \label{entangled_basis}
\end{eqnarray}
which are related to $\ket{\Psi}$ in the same way the Bell states are related to $\ket{\Phi_0}$. The scheme is therefore a verification measurement of an operator possessing these states, along with $\ket{\Psi}$, as eigenstates. When $\theta=\pi/2$ the scheme is the demolition verification measurement of the Bell operator already presented in \fir{frameworkfig}(b). We shall consider shortly in \secr{operator_measure} the more complicated task of simultaneously verifying an arbitrary set of eigenstates.

\subsection{Bipartite multi-qubit states}
\label{multiqubit_verify}
The verification scheme for two-qubit states can be generalized for any state $\ket{\Psi} \in \mathcal{H}_A\otimes\mathcal{H}_B$ split between two parties $A$ and $B$, where $\mathcal{H}_A = (\mathbbm{C}^2)^{\otimes\, v}$ and $\mathcal{H}_B = (\mathbbm{C}^2)^{\otimes\, w}$ are tensor-products of qubits. Again the scheme operates by performing a nonlocal unitary $U$ which maps $\ket{\Psi}$ to a locally measurable state as $U\ket{\Psi} \mapsto \ket{0,0,\cdots,0}$, modulo Pauli distortions. As with two-qubits the scheme focuses on the Schmidt decomposition of $\ket{\Psi}$ which now takes the form
\begin{eqnarray}
\ket{\Psi} &=&
\sum_{\alpha=1}^\chi\lambda_\alpha\ket{\phi_\alpha}_A\ket{\phi_\alpha}_B,
\nonumber
\end{eqnarray}
where $\chi \leq \min(2^v,2^w)$ is the Schmidt rank designating the number of non-zero $\lambda_\alpha$ Schmidt coefficients satisfying  $\sum_{\alpha}\lambda_{\alpha}^2 = 1$, and $\kets{\phi_\alpha}$ are Alice's $(A)$ or Bob's $(B)$ local Schmidt states. Before starting the verification scheme Alice and Bob use this canonical form for the target state to determine local unitaries $V^\dagger_A$ and $V^\dagger_B$ which can be applied to their $v$- and $w$-qubit subsystems, respectively, to map their local Schmidt states into the computational basis. For either party this takes the form
\begin{eqnarray}
V^\dagger\ket{\phi_\alpha} &=& \ket{\vec{\alpha},0,\cdots,0},
\nonumber
\end{eqnarray}
where $\vec{\alpha}$ is a $d = \ceil{\log_2(\chi)}$ dimensional binary vector representing the integer index $\alpha$ and $\ket{\vec{\alpha}}$ is a $d$-fold tensor product of the $\sigma_z$ eigenstates $\ket{0}$ and $\ket{1}$. The action of the $V$'s on the orthogonal complement to the subspace spanned by the local Schmidt states $\ket{\phi_\alpha}$ can be defined arbitrarily. The resulting state $\ket{\psi} = V^\dagger_A\otimes V^\dagger_B\ket{\Psi}$ is then entirely contained in the smallest possible subspace of the original $(v+w)$-qubit system composed of two equal-sized $d$-qubit subsystems at $A$ and $B$. Once this initial compression is performed the remaining $v-d$ and $w-d$ qubits at Alice and Bob's location containing (some of) the orthogonal complement to $\ket{\psi}$ can be immediately measured in the computational basis. Any outcome other than $\ket{0}$ for each qubit indicates an immediate ``no'' result. 

Having mapped the target state $\ket{\Psi}$ to $\ket{\psi}$ the scheme then continues by implementing the inverse of the circuit which locally constructs $\ket{\psi}$ from the $2d$-qubit initial state $\ket{0,0,\cdots,0}$. Specifically this construction circuit begins by creating a superposition state $\ket{\Lambda_d}$ on the first $d$ qubits (generalizing $\ket{\Lambda_1}$ from earlier) of the form
\begin{eqnarray}
\ket{\Lambda_d} &=& \sum_{\vec{x}}\lambda_{\vec{x}}\ket{\vec{x}}
\nonumber
\end{eqnarray}
where $\lambda_{\vec{x}}$ are the real Schmidt coefficients of $\ket{\Psi}$ indexed by the $d$-dimensional binary vector $\vec{x}$ and appropriately padded with zeros if necessary. This type of superposition state can be formed by a cascade $F^0_1 F^1_2 F^2_3 \cdots F^{d-1}_d$ of so-called uniformly controlled rotations (see \ref{uniform} and \cite{Mottonen} for details) about the $y$-axis acting on the first set of $d$ qubits. Once the state $\ket{\Lambda_d}\otimes\ket{0,\cdots,0}$ has been generated a {\em staircase} sequence of CNOT gates are applied between pairs of qubits from the first set of $d$ and the second set of $d$ (see \fir{multi_state_circuit}). This then constructs the canonical Schmidt form for the state $\ket{\psi}$ as
\begin{eqnarray}
\ket{\psi} &=&\sum_{\vec{x}}\lambda_{\vec{x}}\ket{\vec{x}}
\otimes\ket{\vec{x}}. \nonumber
\end{eqnarray}
A more detailed description of this circuit is given in \ref{superposition_state} where it is shown explicitly for $d=4$ qubits in \fir{multi_state_circuit}.

\begin{figure}[t]
\begin{center}
\includegraphics[width=10.8cm]{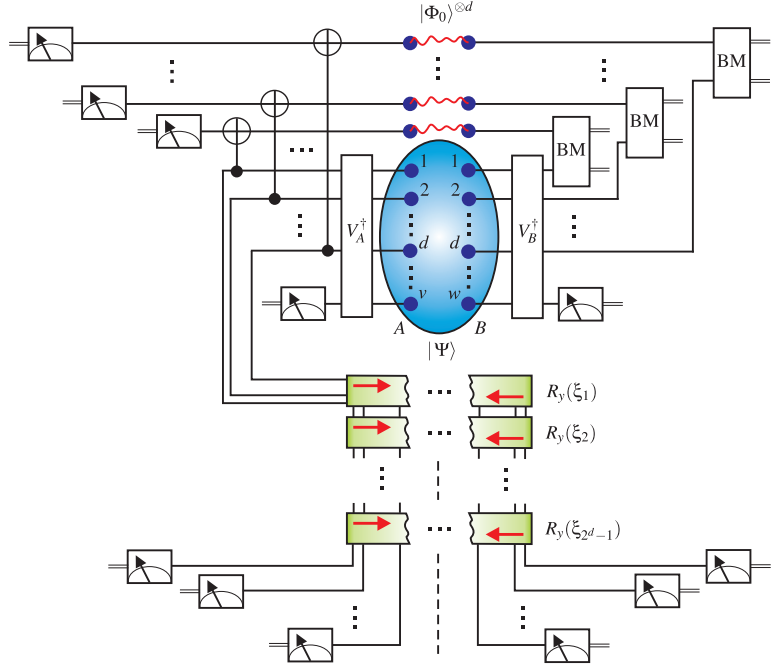}
\caption{The nonlocal verification scheme for a state $\ket{\Psi} \in (\mathbbm{C}^2)^{\otimes v} \otimes (\mathbbm{C}^2)^{\otimes w}$. After performing local unitaries $V_A^\dagger$ and $V_B^\dagger$ which map the target state $\ket{\Psi}$ to a $2d$-qubit state $\ket{\psi}$ and measuring out the orthogonal complement, Bob teleports his $d$ qubits to Alice. Following the inverse of the circuit in \fir{multi_state_circuit} Alice performs a sequence of CNOT gates between her $d$ qubits and those received from Bob, with the latter being measured immediately afterwards. Alice then inverts the sequence of uniformly controlled rotations in the $y$-axis which produce $\ket{\Lambda_d}$ via $2^d-1$ concatenated $d$-qubit rotation chains. Note that although the concatenated rotation chains are drawn sequentially they should be understood as forming a massively recursive structure. The output from the final chain is then measured. We have also ignored here the optimisation that successive sets of rotation chains act on smaller number of qubits due to the structure of the circuit in \fir{multi_state_circuit}.}\label{multi_state_scheme}
\end{center}
\end{figure}

Given this construction circuit the verification scheme proceeds with Bob teleporting his $d$ qubits to Alice. She then implements the sequence of CNOT gates locally on the $2d$ qubits in her possession. Since this part of the circuit is a stabiliser it is guaranteed to succeed but will propagate Pauli distortions originally confined to the ancilla qubits receiving Bob's half of the system to Alice's half. This leaves a state of the form $\sigma_{\bf j}\ket{\Lambda_d}\otimes\ket{0,\cdots,0}$ in Alice's possession, but with only Bob knowing ${\bf j}$. Since the CNOT's have successfully disentangled the two halves the qubits originating from Bob are now in a product state in the computational basis and can be measured immediately. Alice is now left with her $d$ qubits which require the final sequence of uniformly controlled rotations to be applied. The decomposition of the cascading sequence of uniformly controlled rotation into Pauli rotations requires $2^d-1$ distinct gates (\fir{uniform_rotations}(b) shows this decomposition for $F^2_3$) which, as expected, is identical to the number of independent rotation angles defining $\ket{\Lambda_d}$. The scheme then implements these rotations by concatenating rotation chains. The complete nonlocal verification scheme for $\ket{\Psi}$ is shown in \fir{multi_state_scheme}.

Combining the scaling in the number of rotations with that of the average consumption $\av{c_n}$ for concatenated rotations in \eqr{growth} yields an exponential of an exponential scaling
\begin{eqnarray}
\av{e} &=& Cd\phi^{2^d-1}~\textrm{ebits}, \nonumber
\end{eqnarray}
with the minimum number of qubits $d$ required to contain the Schmidt rank of the target state $\ket{\Psi}$. While this consumption lacks any dependence on the values of the angular parameters (excluding binary angles), it does depend on the entanglement in $\ket{\Psi}$ as measured by the Schmidt rank. Note that the choice of unitary $U$ which can implement a verification of a single state $\ket{\Psi}$ is not unique. However, the choice made in this scheme is unique in the sense that it is defined only by the nonlocal parameters of the target state itself and is therefore the most economical. Also, similar to the two-qubit case, this state verification scheme is also a verification measurement of a special operator whose complete set of eigenstates spanning the orthogonal complement also happen to be mapped to locally measurable states. 

\section{Instantaneous measurements of nonlocal operators}
\label{operator_measure}
We now generalize the measurement schemes introduced so far to perform a simultaneous demolition verification of each of the non-degenerate eigenstates of an arbitrary nonlocal observable $O$. Our strategy is again to implement a nonlocal unitary $U$ which maps each eigenstate of $O$ into a different computational basis state which is then locally measurable. Unlike the state verification scheme, which is already a special class of operator measurement, here we are interested in complete generality.

\subsection{Two-qubit observables}
\label{twoqubit_measure}
Before outlining a scheme for the most general case we first describe some schemes for special classes of eigenstates for two-qubits. A particularly interesting class of observables are those with a {\em twisted} eigenbasis,
\begin{eqnarray}
\ket{\Psi_0} = \ket{0}_A\ket{0}_B, \nonumber \\
\ket{\Psi_1} = \ket{0}_A\ket{1}_B, \nonumber \\
\ket{\Psi_2} = \ket{1}_A\left[ \sin\left(\half\theta\right)\ket{0}_B + e^{i\varphi}\cos\left(\half\theta\right)\ket{1}_B\right], \nonumber \\
\ket{\Psi_3} = \ket{1}_A\left[ \cos\left(\half\theta\right)\ket{0}_B - e^{i\varphi}\sin\left(\half\theta\right)\ket{1}_B\right]. \label{twisted_basis}
\end{eqnarray}
Despite these eigenstates being product states it has been shown that if an ideal measurements of this basis were possible it would allow violations of causality~\cite{Groisman01,Vaidman03}. Unlike the direct (or untwisted) product basis a verification measurement of the twisted product basis requires entanglement~\cite{Vaidman03}. As seen in \fir{two_qubit_observe}(a) the circuit which generates this basis locally can straightforwardly yield the nonlocal measurement scheme in \fir{two_qubit_observe}(b) which utilizes just one single-qubit rotation chain. The average entanglement consumption for this basis is dependent on the eigenstate requiring $\av{e}=4$ ebits for $\ket{\Psi_0}$ and $\ket{\Psi_1}$ (where no rotation is needed), or $\av{e}=6$ ebits for $\ket{\Psi_2}$ and $\ket{\Psi_3}$. In this way the measurement of the twisted basis is very similar to the eigenbasis in \eqr{entangled_basis} encountered for state verification. There the eigenbasis was composed of equally but partially entangled eigenstates and needed $\av{e}=6$ ebits for all eigenstates. The consumption for entangled eigenstates, however, grows quickly even with a slight generalization. For instance adding an identical relative phase $e^{i\varphi}$ to all of the basis states in \eqr{entangled_basis} necessitates the concatenation of two single-qubit rotation chains (first for the $z$-axis and second for the $y$-axis) and elevates the consumption to $\av{e}=21$ ebits. Generalizing further gives an eigenbasis composed of partially but unequally entangled eigenstates with differing relative phases, as
\begin{eqnarray}
\ket{\Psi_0} = \sin\left(\half\theta_1\right)\ket{0}_A\ket{0}_B + \cos\left(\half\theta_1\right)e^{i\varphi_1}\ket{1}_A\ket{1}_B, \nonumber \\
\ket{\Psi_1} = \cos\left(\half\theta_1\right)\ket{0}_A\ket{0}_B - \sin\left(\half\theta_1\right)e^{i\varphi_1}\ket{1}_A\ket{1}_B, \nonumber \\
\ket{\Psi_2} = \sin\left(\half\theta_2\right)\ket{0}_A\ket{1}_B + \cos\left(\half\theta_2\right)e^{i\varphi_2}\ket{1}_A\ket{0}_B, \nonumber \\
\ket{\Psi_3} = \cos\left(\half\theta_2\right)\ket{0}_A\ket{1}_B - \sin\left(\half\theta_2\right)e^{i\varphi_2}\ket{1}_A\ket{0}_B, \nonumber
\end{eqnarray}
described by 4 real parameters. From the local preparation circuit shown in \fir{two_qubit_observe}(c), which contains two uniformly controlled rotations, a total of 4 Pauli rotations are required (one for each parameter). The corresponding nonlocal measurement scheme is shown in  \fir{two_qubit_observe}(d). The first three rotations require two-qubit chains, while the last acts only on the second qubit and so can be reduced to a single-qubit chain. Following the calculation in \ref{consumption_calc} the average entanglement consumption for the measurement of this eigenbasis is $\av{e}=224$ ebits.

\begin{figure}[t]
\begin{center}
\includegraphics[width=10.8cm]{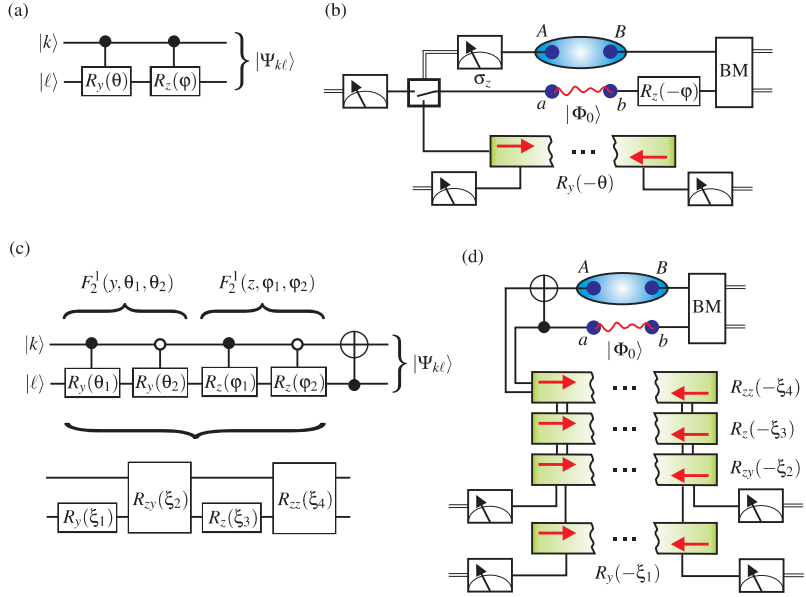}
\caption{(a) A local circuit which constructs the twisted basis set $\ket{\Psi_{k\ell}}$ from the computational basis $\ket{k}\ket{\ell}$. Here the indices $k$ and $\ell$ are bits which together as $k\ell$ are a binary representation of the $\{0,1,2,3\}$ index used in \eqr{twisted_basis}. Firstly a controlled rotation $R_y(\theta)$ is applied to form the twist and then another rotation $R_z(\varphi)$ is performed to introduce the phase. (b) The nonlocal measurement scheme which performs the inverse of (a). The phase can be removed locally by Bob while the final controlled rotation $R_y(-\theta)$ can be replaced by a classical control. If a rotation is required it is implemented by a rotation chain with the output being measured in the $z$-axis. (c) The local circuit which constructs a general partially entangled basis set from $\ket{k}\ket{\ell}$. Dependent on the state $\ket{k}$ the second qubit is rotated about the $y$- and $z$-axis by different angles according to two uniformly controlled rotations, followed by a CNOT gate which entangles them. The uniformly controlled rotations can be decomposed into the sequence of Pauli rotations shown. (d) The nonlocal measurement scheme which performs the inverse of (c). A concatenated sequence of two-qubit rotation chains is applied implementing the inverse of the Pauli rotation decomposition in (c). Note that the final rotation is applied to one qubit only.}\label{two_qubit_observe}
\end{center}
\end{figure}

To devise a scheme to deal with the most general eigenbasis we require a circuit composed only of Pauli rotations, each of which can be handled with rotation chains, that can build a general SU(4) unitary $U$. For two-qubits this can be accomplished by using the so-called Cartan decomposition~\cite{Cirac, Dur} of an SU(4) unitary as
\begin{eqnarray}
U &=& \left(V_A\otimes
V_B\right)e^{\frac{i}{2}\xi_1\sigma_1\otimes\sigma_1}e^{\frac{i}{2}\xi_2\sigma_2\otimes\sigma_2}e^{\frac{i}{2}\xi_3\sigma_3\otimes\sigma_3}\left(W_A\otimes
W_B\right)\nonumber,
\end{eqnarray}
where $V_A$, $V_B$, $W_A$ and $W_B$ are single qubit SU(2) gates, and $\pi/2 \geq \xi_1 \geq \xi_2 \geq |\xi_3| \geq 0$. The Cartan decomposition has been extremely popular in recent work~\cite{Kraus, Zhang, Makhlin} on quantum circuits since it beautifully exposes the nonlocal content of any two-qubit unitary. Rather than needing to consider all 15 real parameters the classification of two-qubit unitaries reduces to the three coordinates $(\xi_1,\xi_2,\xi_3)$ and allows the set of locally inequivalent gates to be characterized geometrically as points within a tetrahedron~\cite{Zhang}. In the context of nonlocal measurements the first pair of unitaries $W_A$ and $W_B$ can be trivially applied by each party locally before the start of the scheme. If the last pair of single-qubit unitaries are then decomposed as a sequence of rotations $V = R_z(\alpha)R_y(\beta)R_z(\gamma)$, we see that the $U$ is expressed entirely in terms of Pauli rotations. Furthermore since our final measurement after $U$ will be in the $z$-axis the latter $R_z$ rotation for either of the local $V$ unitaries is not necessary. This leaves 7 real parameters relevant for the nonlocal measurement.

To compute the average entanglement consumption in this most general case we perform one optimisation. Rather than simply concatenating 7 two-qubit rotation chains (which would consume $2\av{c_7} + 1 = 4719$ ebits on average), we instead split up the qubits after the three nonlocal gates and perform the final two single-qubit rotations on them separately and simultaneously\footnote{Splitting the qubits up can only be done once they never need to interact again. Once separated the qubits progress along different pathways through the scheme and no party knows precisely where the actual pair are located.}. A simple modification of the calculation in \ref{consumption_calc} shows that this splitting gives a consumption equivalent to 5 two-qubit rotation chains and so the average entanglement consumption for the most general two-qubit observable is $2\av{c_5} + 1 = 787$ ebits. Finally, recall from \secr{consumption} that the average consumptions quoted above are upper-bounds to those that would be attained if the angles involved were binary. For example the twisted basis measurement instead consumes at most an average of 3 ebits if $\theta = (2m-1)\pi/8$, where $m$ is an integer.

\subsection{Bipartite multi-qubit observables}
\label{multiqubit_measure}
The situation for mapping the eigenstates of a nonlocal $d$-qubit observable to the computational basis, is less clear due to the lack of an optimal quantum circuit construction for arbitrary SU$(2^d)$ unitaries. On general grounds an exponential number of one-parameter Pauli rotations are expected to be required since an SU$(2^d)$ unitary is defined by $4^d-1$ reals. However, as the two-qubit case illustrates, not all of these parameters are relevant for nonlocal measurements. Recent work \cite{Mottonen} on quantum circuits allows us to identify (although not optimally) some of these redundant local parameters and moreover provides an explicit construction of such a circuit decomposition in terms of Pauli rotations. By exploiting a cosine-sine decomposition recursively a circuit composed only of uniformly controlled rotations was devised in \cite{Mottonen}. So far for $d \geq 4$ qubits this construction represents the most efficient circuit decomposition in terms of the number of CNOT and elementary single-qubit gates needed. For our purposes the important aspects of this construction are that an SU$(2^d)$ unitary can be formed from a circuit of $2^{d+1} -2$ uniformly controlled rotations $F^d_{d-1}$, alternating between the $y$- and $z$-axis, followed by a cascade of $d$ uniformly controlled rotations $F^0_1 F^1_2 \cdots F^d_{d-1}$ involving sequentially decreasing numbers of qubits and all in the $z$-axis . To illustrate this a complete decomposition~\cite{Mottonen} of a $d=3$ qubit gate is shown in \fir{three_qubit_decomp}. If this type of decomposition is used in a nonlocal measurement scheme then the final cascade (shaded in the example in \fir{three_qubit_decomp}) can be ignored since all qubit measurements terminating the circuit are performed in the computational basis. As shown in \ref{superposition_state} each $F_{d-1}$ gate requires $2^{d-1}$ Pauli rotations, equal to the number of reals defining it. Thus using this circuit construction $4^d + 2^d$ concatenated $d$-qubit rotation chains are needed to implement an arbitrary SU$(2^d)$ unitary. An exponential of an exponential scaling with the number of qubits $d$ again arises for the average entanglement consumption $\av{e}$ for a nonlocal measurement of a bipartite multi-qubit observable.

\begin{figure}[t]
\begin{center}
\includegraphics[width=10.8cm]{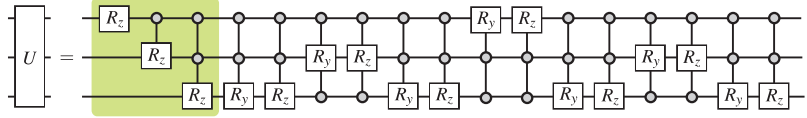}
\caption{The quantum circuit for an arbitrary $d=3$ qubit SU$(2^d)$ unitary $U$ in terms of $17$ uniformly controlled rotation gates (see~\cite{Mottonen} and \ref{uniform} for more details on these gates). The first $2^{d+1}-2=14$ gates alternate between rotations in the $z$- and $y$-axis. The last $d$ gates, which are shaded, are all in the $z$-axis and for nonlocal measurements where $U$ is to map our desired eigenstates to the direct product basis, this final cascade of gates can be ignored.}\label{three_qubit_decomp}
\end{center}
\end{figure}

\section{Conclusions}
\label{conclusion}
In this work we have studied in detail the average entanglement consumption for both nonlocal state verification and operator measurements. The approach applied was similar to that of earlier work~\cite{Vaidman03}  where teleportation was employed to first localise the system and then used in a multi-round protocol to implement the mapping $U$ from a general set of states into a locally measurable set. The central advancement here is that in contrast to previous schemes~\cite{Groisman,Vaidman03} this can be done by consuming only a finite amount of entanglement on average, even in the most general cases, while continuing to succeed with certainty. The reason for this is that the application of $U$ is broken up into a sequence of Pauli rotations $R_{\bf j}(\theta)$. By expressing teleportation in terms of Pauli distortions a decomposition of this type has the privileged feature that distortions at each step either leave the operation $R_{\bf j}(\theta)$ intact or produce only one type of failure, namely $R_{\bf j}(-\theta)$. This enabled us to construct a rotation chain scheme with a bipartite termination condition that applies $R_{\bf j}(\theta)$ with certainty and consumes only a finite amount of the initial entanglement on average. Moreover we showed how rotation chains can be concatenated forming a recursive structure that permits arbitrarily complex sequences of them to be applied while retaining a finite average consumption overall. As an aside this result also shows that bipartite distributed quantum computation can be performed instantaneously with only a finite average entanglement consumption. Interesting comparisons with distributed cluster state one-way quantum computation could be made~\cite{Raussendorf01,Raussendorf03}.

Despite the finiteness of the entanglement consumption its growth is found to display an extremely unfavourable exponential of an exponential scaling with either the Schmidt rank of the state to be verified, or size of the system on which the nonlocal observable acts. While this scaling is scheme dependent there is good reason to believe that it is fundamental to the underlying problem. Indeed, both the complexity of constructing circuit decompositions of a general unitary and the recursive protocol required to overcome the no-signalling constraints individually display exponential scaling. Whether causality forces any conceivable nonlocal measurement scheme to have this combination of scalings is an open problem.

Our aim here has been to prove that general nonlocal measurements can be accomplished with certainty while consuming on average only a finite amount of entanglement. While achieving this the resulting scheme has not been proven to be optimal. Specifically, the consumption in our schemes have no dependence on the actual value of the various rotation angles which appear, beyond the special case of binary angles. Instead the consumption is always averaged over integer units of ebits and the resulting measure of complexity of the required unitary is coarse-grained to simply counting the number of non-trivial rotation angles specifying it. It is possible that a more efficient scheme can be devised where the entangled resources are qubit pairs which are partially entangled, in a way that is linked to the rotation angles, thereby providing a tailored resource and an angle dependent entanglement consumption even for continuous angles.

Another important deficiency of the schemes presented is that they do not yet represent a practical deterministic measurement procedure due to the infinite amount of entanglement that needs to be initially distributed. Here a finite average consumption arises because we have introduced termination conditions for both parties. The requirement for an infinite amount of initial distributed resources appears to be of a different origin, namely the continuous real parameters which appear in the problem. An important exception to this was shown to occur for angles that are binary fractions of $\pi$ where only a finite amount of initial entanglement is needed~\cite{Groisman}. The measurement of the Bell operator is an extreme example of this. Using this result we considered the experimentally relevant case where arbitrary rotation angles are discretised to binary angles. We showed that this results in nonlocal measurement, which is still certain to succeed, but requires only finite initial resources. The resulting measurement performed is an approximation to the exact one and we bounded the error of this procedure. Although not proven it appears unlikely that an exact protocol exists for the most general measurement which succeeds with certainty and requires only a finite amount of initial entanglement. Finally, unlike Vaidman's scheme~\cite{Vaidman03} and stabiliser measurements our rotation chain methods do not easily generalise to more than two-parties and so an interesting open problem is whether all multi-party nonlocal measurements can be done with a finite average entanglement consumption.

\ack SRC and DJ thank the National Research Foundation and the Ministry of Education of Singapore for support. DJ acknowledges support from the ESF program EuroQUAM (EPSRC grant EP/E041612/1), the EPSRC (UK) through the QIP IRC (GR/S82176/01), and the European Commission under the Marie Curie programme through QIPEST. AJC thanks Keble College, Oxford for funding.

\appendix

\section{Computing the average entanglement consumption}
\label{consumption_calc}
In this section the calculation of the average entanglement consumption for the rotation chains used in our scheme is described. We first calculate the average number of channels (complete teleportations of the system) required for the implementation of a single rotation. Using this result and the recursive structure of the scheme we then calculate the average consumption for two rotations, and finally generalise this to an arbitrary number of rotations concatenated together.

\subsection{A single rotation chain}
Following the discussion in \secr{rotation_chain} we consider a chain where Alice possesses the entire system initially and begins the protocol as in \fir{rotation_chain_fig}. The probability that Alice terminates at her $q$th step while Bob terminates at his $p$th step is given by $(\half)^{p+q}$. The consumption of channels is governed by the last party to terminate and denoted by $c_1$.  If Alice terminates last and at her $q$th opportunity then the total number of channels used will be $c_1 = 2q$ and so even. Likewise Bob terminating last at his $p$th opportunity gives a consumption $c_1 = 2p-1$ and is odd. Note that a rotation chain has a minimum consumption of 2 channels and consequently for Bob to terminate last we require $p \geq 2$. The average number of teleportations $\av{c_{1}}$ is then easily calculated as a sum of the case where Alice exits the chain first and when Bob exits the chain first, as
\begin{eqnarray}
\av{c_{1}} &=& \sum_{q=1}^\infty \sum_{p=1}^q \left(\frac{1}{2}\right)^{q+p} 2q + \sum_{p=2}^\infty \sum_{q=1}^{p-1} \left(\frac{1}{2}\right)^{q+p} (2p-1) = 5\nonumber.
\end{eqnarray}
In order to calculate the average consumption when two rotation chains are concatenated we need to introduce two more average consumptions of a single chain. Specifically we define $\av{a_1}$ as the average consumption when only the initiating party is actually performing any actions on the chain, and likewise $\av{b_1}$ for the case where only the receiving party is performing any actions on the chain. These are readily computed as
\begin{eqnarray}
\av{a_{1}} &=& \sum_{q=1}^\infty \left(\frac{1}{2}\right)^{q} 2q= 4, \quad \textrm{and} \quad \av{b_{1}} = \sum_{p=1}^\infty \left(\frac{1}{2}\right)^{p} (2p-1)= 3\nonumber.
\end{eqnarray}
In the next two subsections we shall generalize these quantities to $\av{a_n}$, $\av{b_n}$ and $\av{c_n}$ to designate the corresponding average consumptions for $n$ chains concatenated where only the initiating party, only the receiving party, or both parties are performing actions from the start, respectively.

\subsection{Two rotations chains concatenated}
As with a single chain the consumption for two concatenated rotation chains breaks into two cases depending on whether Alice or Bob exit the first chain last.  In \fir{rotation_concat_fig} and \fir{concat_consump_fig} the latter situation is illustrated with Alice exiting the first chain on the $q$th opportunity, while Bob exits on his $p$th, with $p>q$. Up to her exit point Alice must play the role of the receiving party on all the second chains Bob has available to him at his exit points. Since, for this case, Bob has not used any of these chains Alice consumes $q\av{b_1}$ channels on average through these redundant actions. Similarly Bob must be a receiving party for all $p-1$ of Alice's second chains up to his exit point $p$ consuming $(p-2)\av{b_1} + \av{c_1}$ channels on average, with the $\av{c_1}$ accounting for the fact that one second channel (the $q$th) was used by both parties. At his exit point Bob will consume a further $\av{a_1}$ channels for the second chain which only he acts on. Finally, since Bob exits last (so $p \geq 2$) the consumption of channels in the first chain will be $2p-1$. Performing the analogous counting of channels for the opposite case where Alice exits the first chain last and averaging over all the exit points $p,q$ of the first chain with probabilities $(\half)^{q+p}$ gives 
\begin{eqnarray}
\av{c_{2}} &=& 5 + \av{c_1} + \av{a_1} + 2\av{b_1} = 20 \nonumber.
\end{eqnarray}
It is clear from this that the quantity $\av{r_2} = \av{a_1} + 2\av{b_1}$ represents the cost of recursion within the protocol which in this case doubles the consumption from that expected for two independent rotation chains. We can similarly compute the one-party consumptions for two rotations as $\av{a_2} = 4 + \av{a_1} + 2\av{b_1}$ and $\av{b_2} = 3 + \av{a_1} + \av{b_1}$.

\begin{figure}[t]
\begin{center}
\includegraphics[width=7cm]{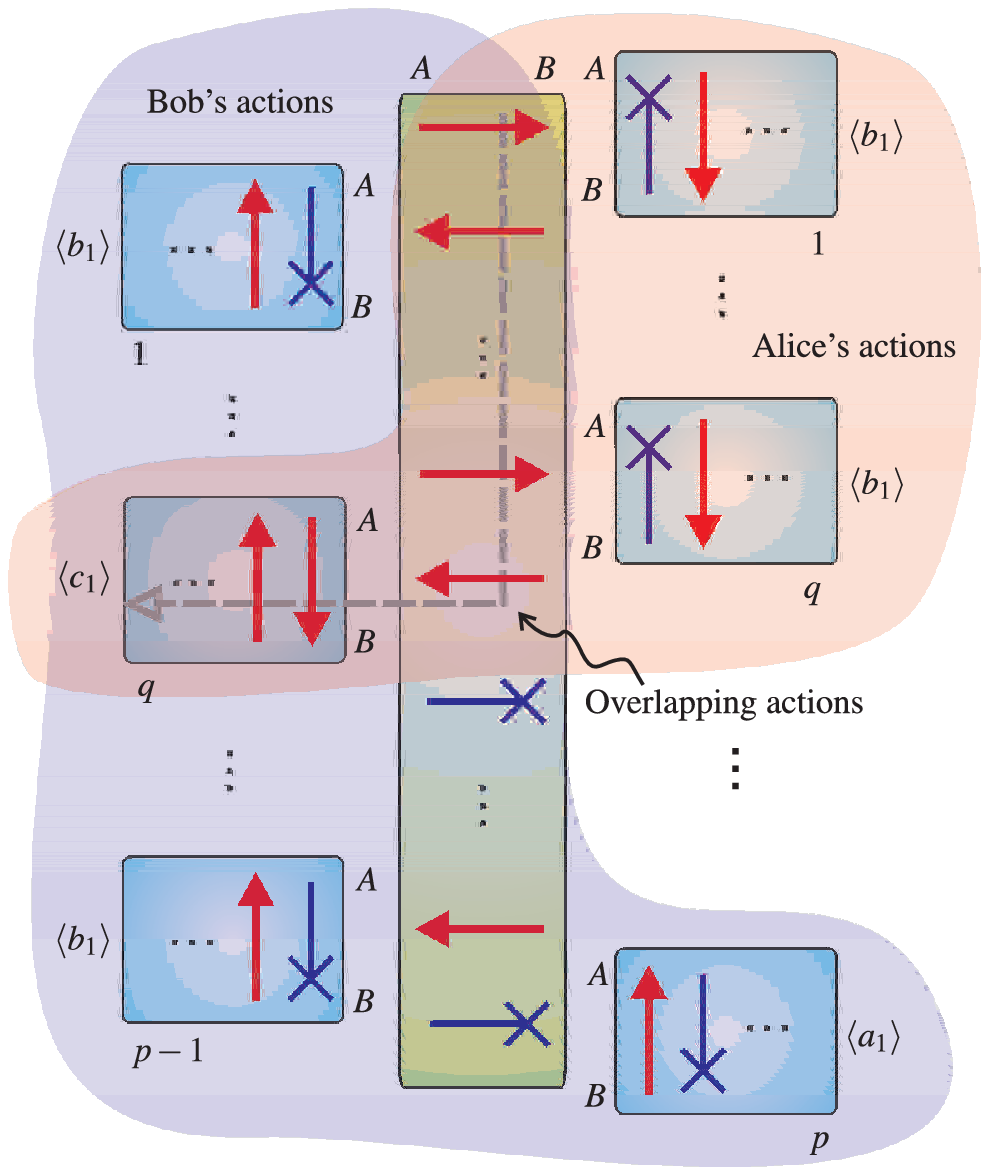}
\caption{Another version of \fir{rotation_concat_fig} but with the consumption quantities $\av{a_1}, \av{b_1}$ and $\av{c_1}$ labelled for the appropriate second chains. The independent actions of Alice and Bob are shaded for the case where Bob exits the first chain last. The dashed ``L''-shaped line indicates the actual path taken by the principal system and lies where their actions overlap.}\label{concat_consump_fig}
\end{center}
\end{figure}

\subsection{Concatenating $n$ rotation chains}
The generalization to $n$ concatenated rotation chains can be computed straightforwardly by using the recursive structure of the protocol. The calculation proceeds in an identical way to two chains except that each second chain itself is now regarded as a sequence of $n-1$ chains. This gives the linked recurrence relations for the component consumptions
\begin{eqnarray}
\av{c_{n}} &=& 5 + \av{c_{n-1}} + \av{a_{n-1}} + 2\av{b_{n-1}}, \nonumber \\
\av{a_n} &=& 4  + \av{a_{n-1}} + 2\av{b_{n-1}}, \nonumber \\
\av{b_n} &=& 3 + \av{a_{n-1}} + \av{b_{n-1}}. \nonumber
\end{eqnarray}
After denoting the recursive consumption as $\av{r_n} = \av{a_{n-1}} + 2\av{b_{n-1}}$ we see that it obeys a closed recurrence relation $\av{r_n} = 3\av{r_{n-1}} - \av{r_{n-2}} - \av{r_{n-3}}$. Given the recursive consumptions $\av{r_1} = 0, \av{r_2} = 10$ and $\av{r_3} = 34$ a closed solution for $\av{r_n}$ can be found as
\begin{eqnarray}
\av{r_{n}} &=& A\phi^n + B\left(\frac{-1}{\phi}\right)^n - 7\nonumber,
\end{eqnarray}
where $A = (3+2\sqrt{2})/2$, $B = (3-2\sqrt{2})/2$, and $\phi = 1 + \sqrt{2}$. For all but the smallest $n$ the recursive consumption $\av{r_n}$ is well approximated by only the first term and so, as might be anticipated, displays a pure exponential growth with $n$. Since the total average consumption is $\av{c_{n}} = 5n +\sum_{k=1}^n \av{r_k}$ it also displays a pure exponential scaling asymptotically as
\begin{eqnarray}
\av{c_{n}} &\sim& A\left(\sum_{k=1}^n \frac{1}{\phi^k}\right)\phi^n \approx C \phi^n \nonumber,
\end{eqnarray}
where $C = (10 + 7\sqrt{2})/4$. As shown in \fir{consumption_fig} this approximation to the exact consumption is already very good once $n>4$.

\section{Uniformly controlled rotations}
\label{uniform}
We make repeated use of a special sequence of multi-qubit controlled rotation gates which, following the nomenclature of \cite{Mottonen}, are called a uniformly controlled rotation. This gate is denoted as $F^k_n({\bf a},\vec{\theta})$ and signifies a $k$-fold controlled rotation of some qubit $n$ about the three-dimensional axis $\bf a$ by one of the $2^k$ different rotation angles contained in $\vec{\theta} = (\theta_1,\theta_2,\cdots,\theta_{2^k})$. The uniformly controlled rotation where qubits $1,\cdots,n-1$ are the controls and qubit $n$ is the target has a matrix representation
\begin{eqnarray}
F^{n-1}_n({\bf a},\vec{\theta}) &=& \left(\begin{array}{ccc}
     R_{\bf a}(\theta_1)& & \\
     & \ddots & \\
     & & R_{\bf a}(\theta_{2^{n-1}})
\end{array}\right). \nonumber
\end{eqnarray}
This gate is motivated by its easily interpreted action, namely it can be seen to implement a different rotation angle on qubit $n$ dependent on each of the $2^{n-1}$ basis configurations of the control qubits. In \fir{uniform_rotations} (a) the circuit defining $F^3_4$ is shown. For our applications we shall exclusively consider rotations in either the $y$-axis $F^k_n(y)$ or $z$-axis $F^k_n(z)$ and we will frequently use a construction which decomposes such $F^k_n$'s into $2^k$ single-parameter Pauli rotations. Specifically for a uniformly controlled rotation $F^{k}_n(y,\vec{\theta})$ this construction involves performing a single qubit rotation $R_y$ on qubit $n$, followed by two-qubit rotations $R_{zy}$ between each of the $k$ control qubits and qubit $n$, followed by three qubit rotations $R_{zzy}$ between every pair of the $k$ control qubits and qubit $n$, and so on until a final rotation $R_{zz\cdots zy}$ is performed involving all the $k$ control qubits and qubit $n$. For this example the Pauli strings for the rotations always specify a $\sigma_y$ on qubit $n$ and $\sigma_z$ on any of the $k$ control qubits. Each of the $2^k$ rotations involves a different rotation angle which itself is a linear combination of the angles in $\vec{\theta}$. A detailed example of this decomposition for a single $F^2_3(y)$ gate is given in \fir{uniform_rotations}(b), while in \fir{two_qubit_observe}(b) a decomposition for the pair of gates $F^1_2(y)F^1_2(z)$ is depicted.

\begin{figure}[t]
\begin{center}
\includegraphics[width=10.8cm]{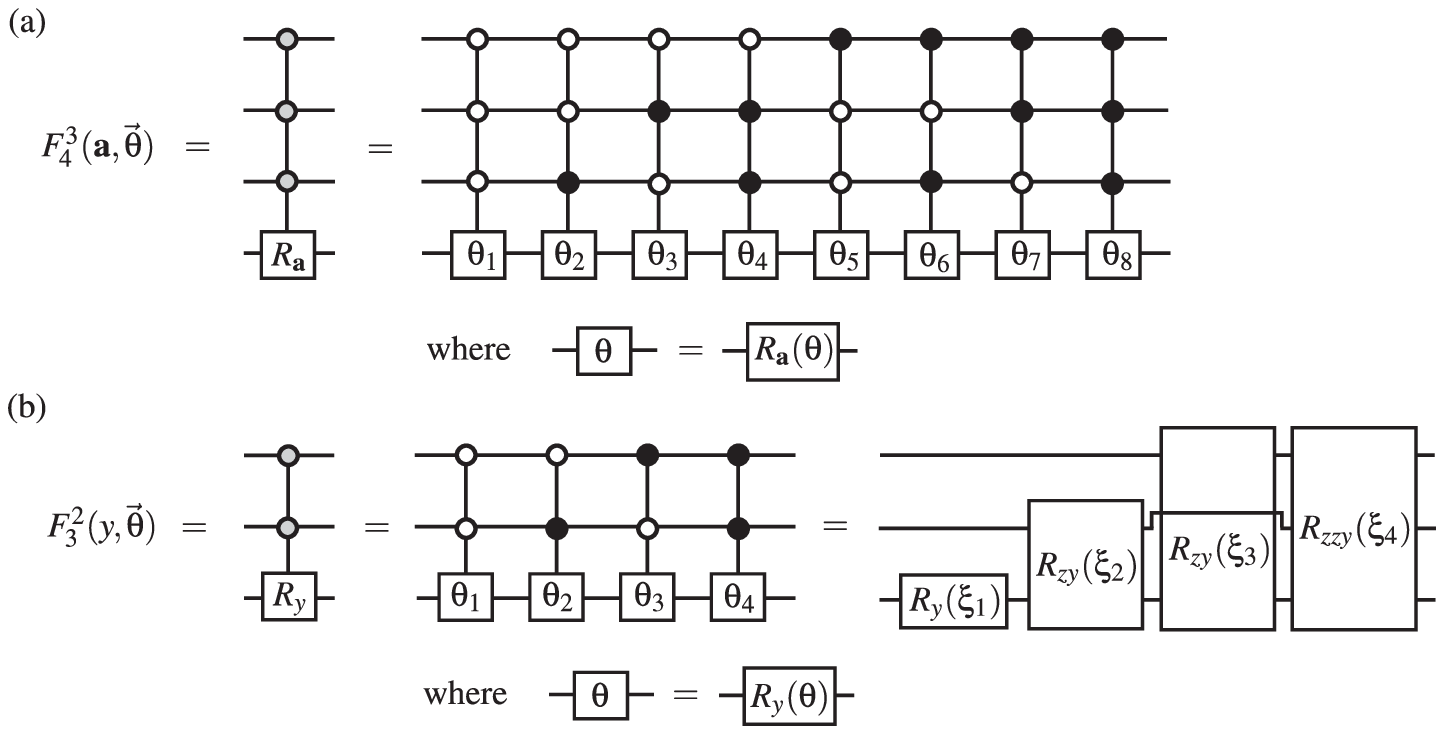}
\caption{(a) The circuit of multi-qubit controlled rotations which constructs the uniformly controlled rotation $F^3_4({\bf a},\vec{\theta})$ about an axis $\bf a$ defined by an eight component vector of angles $\vec{\theta}$. The gate symbol we use for a uniformly controlled rotation is on the left with grey circles. (b) A decomposition in terms of Pauli rotations is shown for a uniformly controlled rotation $F^2_3(y,\vec{\theta})$ about the $y$-axis and defined by a four component vector $\vec{\theta}$. The corresponding Pauli rotation angles are given are related to the four angles in $\vec{\theta}$ as $\xi_1 = -(\theta_1 + \theta_2 + \theta_3 + \theta_4)/8$, $\xi_2 = (\theta_2 -\theta_1 - \theta_3 + \theta_4)/8$, $\xi_3 = (\theta_3 + \theta_4 -\theta_1 - \theta_2)/8$ and $\xi_4 = (\theta_2 + \theta_3 - \theta_1 - \theta_4)/8$. This decomposition readily generalises to uniformly controlled rotations involving larger numbers of qubits.}\label{uniform_rotations}
\end{center}
\end{figure}

\section{Constructing a Schmidt superposition state}
\label{superposition_state}

\begin{figure}[t]
\begin{center}
\includegraphics[width=13cm]{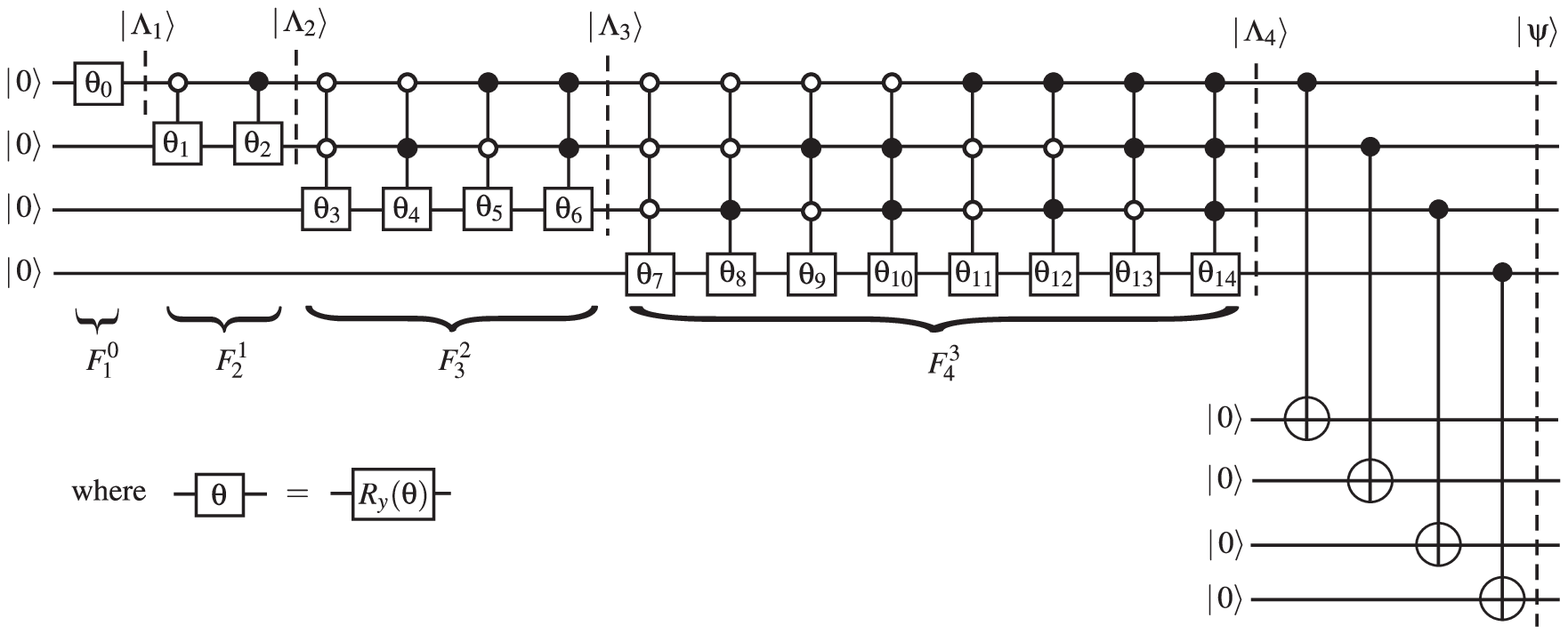}
\caption{The quantum circuit for constructing the state $\ket{\psi}$ composed of 8 qubits. The first part of the circuit constructs the 4 qubit state $\ket{\Lambda_4}$ using a cascade of uniformly controlled rotations. The resulting state $\ket{\Lambda_4}$ is a normalized superposition of each of the $2^4$ computational basis states with real amplitudes parameterized by the 15 angles $\theta_0, \cdots, \theta_{14}$. These angles are chosen to be the Schmidt coefficients of the target state $\ket{\psi}$. The final part of the circuit performs a sequence of CNOT gates between the first 4 qubits and an additional 4 creating the canonical Schmidt form for $\ket{\psi}$. This circuit can be readily generalized to larger numbers of qubits. In particular the structure of the first part of the circuit is based on taking the previous $k-1$ qubits in the state $\ket{\Lambda_{k-1}}$, adding qubit $k$ in the state $\ket{0}$ and performing a further uniformly controlled rotation $F^{k-1}_k$. The resulting $k$ qubits are then expanded to the state $\ket{\Lambda_k}$ and an additional $2^{k-1}$ angles are introduced into its parametrization. See \ref{superposition_state} for more details. }\label{multi_state_circuit}
\end{center}
\end{figure}

For bipartite multi-qubit state verification in \secr{multiqubit_verify} a circuit is required which generates a normalized state in a superposition of all $2^d$ computational basis states with arbitrary real amplitudes. Such a superposition state can be parameterized in terms of $2^d-1$ angles $0\leq \theta_j \leq \pi$ with amplitudes given by
\begin{eqnarray}
\lambda_{\vec{x}} &=&
\cos\left(\half\Theta^{[1]}_{x_1}\right)\cos\left(\half\Theta^{[2]}_{x_1x_2}\right)\cdots\cos\left(\half\Theta^{[d\,]}_{x_1x_2\cdots
x_d}\right), \nonumber
\end{eqnarray}
where the angles $\Theta$ are defined from $\theta$ via
\begin{eqnarray}
\begin{array}{llllllll}
     \Theta^{[1]}_0 = & \theta_0 & & \Theta^{[2]}_{00} = \theta_1 & & \Theta^{[3]}_{000} = \theta_3 & & \cdots \\
     \Theta^{[1]}_1 = & \theta_0 - \pi & & \Theta^{[2]}_{01} = \theta_1 -\pi & & \Theta^{[3]}_{001} = \theta_3 - \pi & & \\
      & & & \Theta^{[2]}_{10} = \theta_2 & & \Theta^{[3]}_{010} = \theta_4 & & \\
      & & & \Theta^{[2]}_{11} = \theta_2 - \pi & & \Theta^{[3]}_{011} = \theta_4 - \pi & & \\
      & & & & & \Theta^{[3]}_{100} = \theta_5 & & \\
      & & & & & \Theta^{[3]}_{101} = \theta_5 - \pi & & \\
      & & & & & \Theta^{[3]}_{110} = \theta_6 & & \\
      & & & & & \Theta^{[3]}_{111} = \theta_6 - \pi. & &
\end{array} \nonumber
\end{eqnarray}
For example, when $d=1$ this reduces to $\lambda_0 = \cos(\half\theta_0)$ and $\lambda_1 = \sin(\half\theta_0)$, while for $d=2$ we have $\lambda_0 = \cos(\half\theta_0)\cos(\half\theta_1)$, $\lambda_1 = \cos(\half\theta_0)\sin(\half\theta_1)$, $\lambda_2 = \sin(\half\theta_0)\cos(\half\theta_2)$, and $\lambda_3 = \sin(\half\theta_0)\sin(\half\theta_2)$. This parametrization of coefficients naturally arises from a sequence of uniformly controlled rotations defined in \ref{uniform}. The construction of a state $\ket{\Lambda_d}$ is then
achieved by a cascade of uniformly controlled rotations, all around the $y$-axis, involving an incrementally increasing subset $1,\cdots,k$ of the $d$ qubits as $F^{k-1}_k$ giving a circuit $F^0_1 F^1_2 F^2_3 \cdots F^{d-1}_d$. In \fir{multi_state_circuit} the circuit building $\ket{\Lambda_4}$ is shown. This figure also shows that as each successive qubit $k$ is added it becomes entangled with the subset of $k-1$ qubits in the state $\ket{\Lambda_{k-1}}$ previously rotated leaving an enlarged total state $\ket{\Lambda_k}$ that is completely defined by the $2^k-1$ independent angles $\theta_j$.

\end{document}